\newcommand*{\ATLASLATEXPATH}{latex/}
\documentclass[cernpreprint,texlive=2016,txfonts,UKenglish]{latex/atlasdoc}
\usepackage{\ATLASLATEXPATH atlaspackage}
\usepackage{\ATLASLATEXPATH atlasbiblatex}
\usepackage{\ATLASLATEXPATH atlasphysics}
\graphicspath{{logos/}{figures/}}

\usepackage{main-defs}
\usepackage{booktabs,siunitx}
\usepackage{dcolumn}
\usepackage{multirow}
\AtlasTitle{\boldmath Measurement of the \ttbar\ production cross-section using $e\mu$ events with $b$-tagged jets  
in $pp$ collisions at $\sqrt{s}=13$\,\TeV\ with the ATLAS detector}

\author{The ATLAS Collaboration}

\AtlasRefCode{TOPQ-2015-09}
\PreprintIdNumber{CERN-PH-2016-088}

\AtlasJournalRef{\PLB 761 (2016) 136}
\AtlasDOI{10.1016/j.physletb.2016.08.019}

\AtlasAbstract{%
This paper describes a measurement of the inclusive top quark pair production 
cross-section (\xtt) with a data sample of 
\intlumi\,\ifb\ of proton--proton collisions at a centre-of-mass energy 
of \sxy, collected in 2015 by the ATLAS detector at the LHC.  This measurement uses events 
with an opposite-charge electron--muon pair in the final state. Jets containing $b$-quarks 
are tagged using an algorithm based on track impact parameters and
reconstructed secondary vertices. The numbers of events 
with exactly one and exactly two $b$-tagged jets are counted and used to
determine simultaneously \xtt\ and the efficiency to reconstruct and
$b$-tag a jet from a top quark decay, thereby
minimising the associated systematic
uncertainties. The cross-section is measured to be:
\begin{center}
\xtt\ = \ttxval\ $\pm$ \ttxstat\ (stat) $\pm$ \ttxsyst\ (syst) $\pm$ \ttxlumi\ (lumi) $\pm$ \ttxebeam\ (beam)~pb,
\end{center}
where the four uncertainties arise from data statistics, experimental and
theoretical systematic effects, the integrated luminosity and the LHC beam energy, 
giving a total relative uncertainty of \ttxrel. The result is consistent
with theoretical QCD calculations at next-to-next-to-leading order.
A fiducial measurement corresponding to the experimental acceptance of the 
leptons is also presented.
}

\hypersetup{pdftitle={ATLAS document},pdfauthor={The ATLAS Collaboration}}

\begin{document}

\maketitle

%\tableofcontents

% List of contributors - print here or after the Bibliography.
%\PrintAtlasContribute{0.30}
%\clearpage

%-------------------------------------------------------------------------------
\section{Introduction}
\label{sec:intro}
%-------------------------------------------------------------------------------
The top quark is the heaviest known fundamental particle, with a mass \mtop\ 
which is much larger than any of the other quarks, and close to the scale
of electroweak symmetry breaking. The study of its production and decay 
properties forms a core part of the LHC physics programme. 
At the LHC, top quarks are primarily produced in quark--antiquark pairs (\ttbar), and the
precise prediction of the corresponding inclusive cross-section is sensitive to the gluon 
parton distribution function (PDF) and the top quark mass, and presents a substantial 
challenge for QCD calculational techniques. 
Physics beyond the Standard Model may also lead to an enhancement of the \ttbar\ production rate.

Calculations of the \ttbar\ production cross-section at hadron colliders are 
available at full next-to-next-to-leading-order (NNLO) accuracy in the strong coupling constant $\alpha_{\textrm S}$, 
including the resummation of next-to-next-to-leading logarithmic (NNLL) soft gluon
terms~\cite{topxtheo1,topxtheo2,topxtheo3,topxtheo4,topxtheo5}. 
In this paper  a reference value of $832^{+40}_{-46}$\,pb at
a centre-of-mass energy of \sxy\ assuming $\mtop=172.5$\,GeV is used,
corresponding to a relative precision of $^{+4.8}_{-5.5}\%$. 
This value was calculated using the \textsc{top++ 2.0} program \cite{toppp}. The combined 
PDF and $\alpha_{\textrm S}$ uncertainties of 
$\pm 35$\,pb were calculated using the PDF4LHC prescription
\cite{pdflhc} with the MSTW2008 68\% CL NNLO \cite{mstwnnlo1,mstwnnlo2}, 
CT10 NNLO \cite{cttenpdf,cttennnlo} and NNPDF2.3 5f FFN \cite{nnpdfffn} 
PDF sets, and added in quadrature to the factorisation and renormalisation scale uncertainty of 
$^{+20}_{-29}$\,pb.
The cross-section at \sxyt\ is predicted to be 3.3 times larger than
the cross-section at \sxvt.

Measurements of \xtt\ have been made at $\sqrt{s}=7$ and $8$\,TeV by both ATLAS~\cite{atopxsec,TOPQ-2013-06,TOPQ-2011-02} and CMS~\cite{CMStop78dil,CMStop8ljet,CMStop7ljet}.
The most precise ATLAS measurements of \xtt\ at these collision energies were made
using events with an opposite-charge isolated electron and muon pair and additional $b$-tagged jets~\cite{atopxsec}. 
This paper documents a measurement of \xtt\ at \sxyt\ using the 
same final state and analysis technique. %, and a sample of \intlumi\,\ifb\ of $pp$ collision data taken in 2015 with 25~ns bunch spacing. 
Wherever possible, the analysis builds on the studies and procedures used in the earlier publication~\cite{atopxsec}. 
A fiducial measurement determining the cross-section in the region corresponding to the experimental 
lepton acceptance is also presented.

The data and Monte Carlo simulation samples are described in Section~\ref{sec:dmc}, 
followed by the object and event selection in Section~\ref{sec:objev} and
the method  for determining the \ttbar\ cross-section in Section~\ref{sec:ext}.
The evaluation of backgrounds is discussed in Section~\ref{sec:back} and 
the systematic uncertainties in Section~\ref{sec:syst}.
Finally, the results and conclusions are given in Section~\ref{sec:result}.

%-------------------------------------------------------------------------------
\section{Data and simulation samples}
\label{sec:dmc}
%-------------------------------------------------------------------------------
The analysis is performed using the full 2015 proton--proton ($pp$) collision data sample at \sxyt\ 
with 25\,ns bunch spacing recorded by the ATLAS detector~\cite{PERF-2007-01,ibl}.
The data correspond to an integrated luminosity of \intlumi\,\ifb\
after requiring stable LHC beams and that all detector subsystems were operational.
Events are required to pass either a single-electron or single-muon trigger,
with thresholds set to be almost fully efficient for leptons with  transverse momentum $\pt>25$\,GeV
passing offline selections. Each event includes the signals 
from on average about 14 additional inelastic $pp$ collisions in the same bunch crossing (known as pile-up).

Monte Carlo simulated event samples are used to optimise the analysis, to compare to the data, and to evaluate signal and background efficiencies and
uncertainties. The samples used in the analysis are summarised in Table~\ref{t:MCSummary}. The main \ttbar\ signal and 
background samples were processed through the ATLAS
detector simulation \cite{SOFT-2010-01} based on GEANT4 \cite{geant4}. Some of the systematic 
uncertainties were studied using alternative \ttbar\ samples processed through a faster simulation 
making use of parameterised showers in the calorimeters \cite{fastsim}. Additional simulated $pp$ collisions generated with
\textsc{Pythia8.186}~\cite{pythia8} were overlaid to model the effects from additional collisions in the same and 
nearby bunch crossings. All simulated events were processed using the
same reconstruction algorithms and analysis chain as the data, and small
corrections were applied to lepton trigger and reconstruction efficiencies and resolutions    
to improve the agreement with the response observed in data.

%%%%%%%%%%%%Signal%%%%%%%%%%%%%%%%%%%%%%%%%%%%%%%%%%
The baseline \ttbar\ simulation sample was produced  at next-to-leading order (NLO) in QCD using the matrix-element generator 
\textsc{Powheg-Box} v2~\cite{powheg,powheg2,powheg3} with CT10 PDFs \cite{cttenpdf}, interfaced to
\textsc{Pythia6} \cite{pythia6} with the Perugia 2012 set of tuned parameters (tune) \cite{perugia} for parton shower, fragmentation and underlying event modelling. 
The \hdamp\ parameter, which gives a cutoff scale
for the first gluon emission, was set to \mtop, a value which was chosen to give good modelling of the \ttbar\ system $p_{\textrm T}$ at \sxwt\ \cite{hdamp}.
The \textsc{EvtGen} \cite{evtgen} package was used to  better simulate the decay of 
heavy-flavour hadrons.

Alternative \ttbar\ simulation samples were generated using \textsc{Powheg}
interfaced to \textsc{Herwig++} \cite{herwigpp}, %(DS 410004, fast simulation),
and \textsc{Madgraph5\_aMC@NLO} \cite{amcnlo} interfaced to \textsc{Herwig++}.
The effects of initial- and final-state radiation were explored using two alternative
\textsc{Powheg\,+\,Pythia6} samples: one with \hdamp\ set to $2\mtop$, the factorisation and renormalisation scale
varied by a factor of 0.5 and using the Perugia 2012 radHi tune, giving more parton shower radiation; % (DS 410001, full simulation),
and a second one with the Perugia 2012 radLo tune, $\hdamp=\mtop$ and  the factorisation and renormalisation scale varied by a factor of 2, giving
less parton shower radiation. % (DS 410002, full simulation). 
The samples were simulated following the recommendations documented in Ref.~\cite{ATL-PHYS-PUB-2016-004}.
The top quark mass was set to 172.5\,GeV in all these simulation samples and the $t\rightarrow Wb$ branching fraction to 100\%.

%%%%%%%%%%%% Backgrounds %%%%%%%%%%%%%%%%%%%%%%%%%%%%%%%%%%%%%%%%%%%%%%%%%%%%
Backgrounds in this measurement are classified into two types: those
with two real prompt leptons from $W$ or $Z$ decays (including those produced 
via leptonic decays of $\tau$-leptons), and those where at least one of the reconstructed 
lepton candidates is `fake', 
{\em i.e.}\ a non-prompt lepton produced from the decay of 
a bottom or charm hadron, an electron arising from a photon conversion, 
a jet misidentified as an electron, or a muon produced from an
in-flight decay of a pion or kaon.
Backgrounds containing two real prompt leptons include single-top production
in association with a $W$ boson ($Wt$), $Z$+jets production with 
$Z\rightarrow\tau\tau\rightarrow e\mu$, and diboson production ($WW$, $WZ$ and
$ZZ$) where both bosons decay leptonically.

The dominant $Wt$ single-top background was modelled using \textsc{Powheg-Box v1\,+\,Pythia6}
with the CT10 PDFs and the Perugia 2012 tune, using the `diagram removal'
generation scheme~\cite{WtDR}.
The $Z$+jets background was modelled using 
\textsc{Sherpa}~2.1.1~\cite{sherpa}: matrix elements (ME) were calculated for up to 
two partons at NLO and four partons at leading order using the \textsc{Comix} \cite{comix} and \textsc{OpenLoops} \cite{openloops} matrix-element 
generators and merged with the \textsc{Sherpa} parton shower (PS) using the \textsc{ME\,+\,PS@NLO} \cite{mepsnlo} prescription; 
the CT10 PDF set was used in conjunction with dedicated parton shower tuning in \textsc{Sherpa}. %developed by the \textsc{ Sherpa} authors.
Diboson production with additional jets was also simulated using \textsc{Sherpa} 2.1.1 and CT10 PDFs as described above;
the  four-lepton final state, the three-lepton final state with two different-flavour leptons, and the two-lepton final state 
were simulated to cover $ZZ$, $ZW$ and $WW$ production, and include off-shell $Z/\gamma^*$ contributions. 
Same-charge $WW$ production from QCD and electroweak processes was included. 
Alternative $Wt$ and diboson simulation samples were generated using \textsc{Powheg\,+\,Herwig++} and \textsc{Powheg\,+\,Pythia8}, respectively, to estimate 
the background modelling uncertainties.

The majority of the background with at least one fake lepton in the selected
sample arises from \ttbar\ production where only one of the $W$ bosons from the top quarks decays leptonically, which was simulated %as for dileptonic \ttbar\ production 
as discussed earlier. Other processes with one real lepton
which can contribute to this background include the $t$-channel single-top
production, modelled using \textsc{Powheg-Box v1\,+\,Pythia6}, and 
$W$+jets with the $W$ decaying to $e\nu$, $\mu\nu$ or
$\tau\nu$ where the $\tau$-lepton subsequently decays leptonically. This
background was modelled using \textsc{Powheg-Box v2\,+\,Pythia8} with the CT10 PDFs.
The small expected contribution from \ttbar\ in association with a $W$ or $Z$ boson to the same-charge $e\mu$ sample used
for background estimation was modelled using \textsc{MadGraph\,+\,Pythia8}~\cite{madgraph}.
Other backgrounds, including processes 
with two misidentified leptons, are negligible.

\begin{table}[tp]
\centering
%\begin{tabular}{l|cc}\hline
\begin{tabular}{l|c|c}
\hline
Process & Generator + Parton Shower & Calculation \\
\hline
\multirow{3}{*}{\ttbar} & \textsc{Powheg-Box v2\,+\,Pythia6}  &\multirow{3}{*}{NLO} \\ 
                        & \textsc{Powheg-Box v2\,+\,Herwig++} & \\ 
                        &  \textsc{Madgraph5\_aMC@NLO\,+\,Herwig++}& \\ 
\hline
\multirow{2}{*}{$Wt$ single top}  & \textsc{Powheg-Box v1\,+\,Pythia6}&\multirow{2}{*}{NLO} \\
  &  \textsc{Powheg-Box v1\,+\,Herwig++} & \\
\hline
$Z$+jets & \textsc{Sherpa} 2.1.1 & NLO (up to two partons)\\
\hline
\multirow{2}{*}{Diboson} & \textsc{Sherpa} 2.1.1 & NLO (up to two partons)\\
 & \textsc{Powheg\,+\,Pythia8} & NLO\\
\hline
$t$-channel single top & \textsc{Powheg-Box v1\,+\,Pythia6} & NLO\\
\hline
$W$+jets & \textsc{Powheg-Box v2\,+\,Pythia8} & NLO\\
\hline
\ttbar\ + $W$/$Z$ & \textsc{MadGraph\,+\,Pythia8} &  LO\\
\hline
\end{tabular}
\caption{\label{t:MCSummary}Summary of Monte Carlo samples used to model the signal and background processes. The 'Calculation' column corresponds 
to the order of the matrix element calculation in the Monte Carlo generator.}
\end{table}

\newcommand{\AtlasCoordFootnote}{%
ATLAS uses a right-handed coordinate system with its origin at the nominal interaction point (IP)
in the centre of the detector and the $z$-axis along the beam pipe.
The $x$-axis points from the IP to the centre of the LHC ring,
and the $y$-axis points upwards.
Cylindrical coordinates $(r,\phi)$ are used in the transverse plane, 
$\phi$ being the azimuthal angle around the $z$-axis.
The pseudorapidity is defined in terms of the polar angle $\theta$ as $\eta = -\ln \tan(\theta/2)$.
Angular distance is measured in units of $\Delta R \equiv \sqrt{(\Delta\eta)^{2} + (\Delta\phi)^{2}}$.}
%-------------------------------------------------------------------------------
\section{Object and event selection}
\label{sec:objev}
%-------------------------------------------------------------------------------
This measurement makes use of reconstructed electrons, muons and $b$-tagged jets.
The object and event selections largely follow those used in the earlier 
publication; 
in particular the same kinematic cuts are used for electrons and jets, and very similar ones are used for muons. 

\textbf{Electron candidates} are reconstructed from an isolated electromagnetic 
calorimeter energy  deposit match\-ed to a track in the inner detector and
passing a medium likelihood-based requirement~\cite{PERF-2013-03,ATLAS-CONF-2014-032}, within the fiducial region of 
transverse energy $\et>25$\,GeV and pseudorapidity\footnote{\AtlasCoordFootnote} $|\eta|<2.47$. Candidates within the
transition region between the barrel and endcap electromagnetic calorimeters,
$1.37<|\eta|<1.52$, are removed. 
The electron candidates must satisfy requirements on the transverse impact parameter significance calculated 
%with respect to the beamline of $\frac{|d_0|}{\sigma_{d_0}}<5$ %$|d_{\textrm 0,sig}|<5$ 
with respect to the beamline of $|d_0|/\sigma_{d_0}<5$ %$|d_{\textrm 0,sig}|<5$ 
and on the longitudinal impact parameter calculated 
with respect to the primary vertex of $|\Delta z_{\textrm 0}$~sin$~\theta|<0.5$~mm. The primary vertex is 
defined as the one with the highest sum of $p_{\textrm T}^2$ of tracks associated to it.
Electrons are required to be isolated using requirements on the calorimeter energy in a cone of size
%$\Delta R$\footnote{$\Delta R=\sqrt{\Delta\eta^2+\Delta\phi^2}$.}$<0.2$ around the electron (excluding the deposit from the 
$\Delta R<0.2$ around the electron (excluding the deposit from the 
electron itself)  divided by the electron $\pt$, and 
on the sum of track $\pt$ in a variable-size cone around the electron 
direction (again excluding the electron track itself).
The track isolation cone size is given by
the smaller of  $\Delta R=10$\,GeV$/\pt(e)$ and $\Delta R=0.2$, 
{\em i.e.}\ a cone which increases in 
size at low \pt\ up to a maximum of 0.2. 
Selection criteria, dependent on \pt\ and $\eta$, are
applied to produce a nominal efficiency of 95\% for electrons from 
$Z\rightarrow ee$ decays with \pt\ of 25\,GeV which rises to 99\% at 60\,GeV. 
The efficiencies
in \ttbar\ events are smaller, due to the increased jet activity.
To prevent double-counting
of electron energy deposits as jets, the closest jet with $\Delta R<0.2$ of 
a reconstructed electron is removed. 
Finally, if the nearest jet surviving
the above selection is within $\Delta R=0.4$ of the electron, the electron
is discarded, to ensure it is sufficiently separated from nearby jet activity.

\textbf{Muon candidates} are reconstructed by combining matching tracks reconstructed
in both the inner detector and muon spectrometer, and are required to satisfy
$\pt>25$\,GeV and $|\eta|<2.4$~\cite{muons13TeV}. 
Muons are also required to be isolated, using requirements similar to those for electrons, with the selection criteria tuned to give similar efficiencies for $Z\rightarrow\mu\mu$ events.
The muon candidates must satisfy the requirements  on the transverse 
 impact parameter significance and on the longitudinal impact parameter 
% of  $\frac{|d_0|}{\sigma_{d_0}}<3$ %$|d_{\textrm 0,sig}|<3$ 
 of  $|d_0|/\sigma_{d_0}<3$ %$|d_{\textrm 0,sig}|<3$ 
and $|\Delta z_{\textrm 0}$~sin$~\theta|<0.5$~mm, respectively.
To reduce the background 
from muons from heavy-flavour decays inside jets, muons are removed if they
are separated from the nearest jet by $\Delta R<0.4$. However, if this jet
has fewer than three associated tracks, the muon is kept and the jet is removed
instead;  this avoids an inefficiency for high-energy muons undergoing 
significant energy loss in the calorimeter.

\textbf{Jets} are reconstructed using the anti-$k_t$ algorithm \cite{antikt,antikt2} 
with radius parameter $R=0.4$, starting from topological clusters of deposited energy in the calorimeters. % ~\cite{Lampl:2008zz}.
Jets are calibrated using an energy- and $\eta$-dependent
simulation-based calibration scheme with corrections derived from data. % in-situ corrections. %based on Run 1 data~\cite{jesxi}, checked with early Run 2 data~\cite{jesrun2}.
No corrections for semileptonic $b$-hadron decays are
applied. Jets are accepted within the fiducial region 
$\pt>25$\,GeV and $|\eta|<2.5$. 
To reduce the contribution from jets 
associated with pile-up, jets with $\pt<50$\,GeV and $|\eta|<2.4$
are required to pass a pile-up rejection veto~\cite{PERF-2014-03}.

Jets are $b$-tagged as likely to contain $b$-hadrons using the
MV2c20 algorithm~\cite{MV2c20}, a multivariate discriminant making use of
track impact parameters and reconstructed secondary vertices and tuned with 
the new detector configuration, {\em i.e.}\ including the Insertable B-Layer detector (IBL)~\cite{ibl}. %\cite{btagcom,btagptrel,btagrun2}. 
Jets are defined as being $b$-tagged if the MV2c20 weight is larger than a threshold value corresponding to approximately 70\% $b$-tagging efficiency for 
$b$-jets in \ttbar\ 
events, although the exact efficiency varies with \pt. In simulation,
the tagging algorithm gives a rejection factor of about 440 
against light-quark and gluon jets, and about 8 against jets originating
from charm quarks. The improvements of a factor of three in the light-quark rejection and of 60\% in the charm-quark rejection compared to 
the $b$-tagging algorithm used in Ref.~\cite{atopxsec} originate from the gain in track impact parameter resolution from the IBL, 
and improvements in the track reconstruction and $b$-tagging algorithms~\cite{MV2c20}.

Events are rejected if the selected electron and muon are separated by $\Delta\phi<0.15$\,rad and
$\Delta\theta<0.15$\,rad, where $\Delta\phi$ and $\Delta\theta$ are the differences
in polar and azimuthal angles between the two leptons. 
This requirement
% is referred to as the muon bremsstrahlung cut, and 
rejects events where a muon undergoes 
significant energy loss in the electromagnetic calorimeter, thus leading to a reconstructed electron candidate.
Events passing the above requirements, and having exactly one
selected electron and one selected muon of opposite electric charge sign (OS),
define the $e\mu$ preselected sample. The corresponding same-sign (SS) sample 
is used in the estimation of background
from events with misidentified leptons.
Events are then further classified into those with exactly one or exactly two 
$b$-tagged jets.

%-------------------------------------------------------------------------------
\section{Extraction of the \ttbar\ cross-section}
\label{sec:ext}
%-------------------------------------------------------------------------------
The \ttbar\ cross-section is measured in the dileptonic $e\mu$ channel, where
one top quark decays as $t\rightarrow Wb\rightarrow e\nu b$ and the other
as $t\rightarrow Wb\rightarrow\mu\nu b$.\footnote{This notation indicates 
the leptonic decay of  both $t$ and $\bar{t}$. Charge-conjugate modes are 
implied unless otherwise stated.} The final states from leptonic $\tau$ decays are also included.  
As in Ref.~\cite{atopxsec}, \xtt\ is determined by counting the numbers
of opposite-sign $e\mu$ events with exactly one ($N_1$) and exactly
two ($N_2$) $b$-tagged jets, ignoring any jets that are not $b$-tagged which may be present,
due to {\em e.g.}\ light-quark or gluon jets from QCD radiation or $b$-jets from
top quark decays which are not $b$-tagged. The two event counts can be expressed as:
\begin{eqnarray}
N_1 & = & L \xtt\ \epsem 2\epsb (1-\cb\epsb) + \nib \nonumber \\
N_2 & = & L \xtt\ \epsem \cb\epsb^2 + \niib \label{e:tags}
\end{eqnarray}
where $L$ is the integrated luminosity of the sample and \epsem\ the 
efficiency for a \ttbar\ event to pass the opposite-sign $e\mu$ preselection. 
The combined probability for a jet from the quark $q$ in the $t\rightarrow Wq$ 
decay to fall within the acceptance of the detector,
be reconstructed as a jet with transverse momentum above the selection 
threshold, and be tagged as a $b$-jet, is denoted by \epsb. 
%Although this 
%quark is almost always a $b$-quark, \epsb\ also accounts for 
%the approximately $0.2\%$ of top quarks that decay to $Ws$ or $Wd$ 
%rather than $Wb$, slightly reducing the effective tagging efficiency.
If the decays of the two top quarks
and the subsequent reconstruction of the two $b$-tagged jets are completely
independent, the probability to tag both $b$-jets \epsbb\ is given by
$\epsbb=\epsb^2$. In practice, small correlations are present for kinematic 
and instrumental reasons, and these are taken into account via
the tagging correlation coefficient $\cb$, defined as $\cb=\epsbb/\epsb^2$ or equivalently
$\cb=4 N^{\ttbar}_{e\mu} N_2^{\ttbar}/(N^{\ttbar}_1+2 N^{\ttbar}_2)^2$,
where $N^{\ttbar}_{e\mu}$ is the number of preselected $e\mu$ \ttbar\ events
and $N^{\ttbar}_1$ and $N^{\ttbar}_2$ are the numbers of events with one and
two $b$-tagged jets.
Background
from sources other than $\ttbar\rightarrow e\mu\nu\nubar\bbbar$
also contributes to the event counts 
$N_1$ and $N_2$, and is given by the background terms
\nib\ and \niib. The preselection efficiency \epsem\ and tagging correlation
\cb\ are taken from \ttbar\ event simulation and are about  $0.83$\% and $1.002$, respectively, and 
the background contributions
\nib\ and \niib\ are estimated using a combination of simulation and 
data-based methods as described in Section~\ref{sec:back}, allowing the two equations~(\ref{e:tags}) to be solved yielding \xtt\ and \epsb\ 
by minimising a likelihood function. % as discussed in more detail in Ref.~\cite{atopxsec}.

In the method to measure the \ttbar\ cross-section outlined above, some of the largest systematic uncertainties come from the use of
simulation to estimate the preselection efficiency \epsem. This efficiency can be factorised into the product
of two terms: $\epsem=\aem\gem$. The acceptance \aem\ represents the fraction of \ttbar\ 
events that have a true $e\mu$ pair within the detector acceptance ($p_{\textrm T}>25$\,GeV and
$|\eta|<2.5$) and it is about 1.4\% (1.2\% excluding $\tau$ decays). 
The term \gem\ represents the ratio of reconstructed \ttbar\ events to \ttbar\ events with a true $e\mu$ pair within the fiducial region, 
where the numerator includes the approximately 2\% of reconstructed \ttbar events where one or both leptons have true $p_{\textrm T}<25$\,GeV.
The fiducial cross-section \xfid\ is defined as $\xfid=\aem\xtt$, avoiding the systematic uncertainties associated with the extrapolation 
from the  measured lepton phase space to the full phase space, and measured following the same technique as in Ref.~\cite{atopxsec}.
The contribution of \ttbar\ events produced in the fiducial region with at least one lepton originating via $W \rightarrow \tau \rightarrow l$ decay is estimated from simulation to be 12.2~$\pm$~0.1\%.

A total of 30879 data events passed the $e\mu$ opposite-sign preselection.
Table~\ref{t:evtsum} shows the number of events with one and two $b$-tagged
jets, together with the estimates of non-\ttbar\ background and their 
systematic uncertainties discussed below. The ratio of $b$-tagged events 
to preselected events (before $b$-tagging) is higher for $13$\,TeV than at $7$ and $8$\,TeV 
due to the larger increase of the \ttbar\ cross-section with $\sqrt{s}$ compared with the 
Z+jets and diboson background cross-sections.
In simulation, the sample with one $b$-tagged jet is expected to be about 89\% pure in
\ttbar\ events, with the dominant background originating from $Wt$ single-top production, 
and smaller contributions from events with misidentified
leptons, $Z$+jets and dibosons.
The sample with two $b$-tagged jets is expected to be about 96\% pure in 
\ttbar\ events, with $Wt$ production again being the dominant background.
% TAGTBL event counts updated 10/02/2016
\begin{table}[tp]
\centering
%\begin{tabular}{l|cc}\hline
\begin{tabular}{l| D{,} {\pm} {-1} D{,} {\pm} {-1}}
\hline
Event counts & \multicolumn{1}{c}{$N_1$} & \multicolumn{1}{c}{$N_2$} \\
\hline

Data & \multicolumn{1}{c}{11958} & \multicolumn{1}{c}{7069}  \\
\hline
Single top                                    &  1140~,~100 & 221 ~,~ 68\\
Diboson                                       &  34  ~,~11  & 1   ~,~ 0\\
$Z(\rightarrow\tau\tau\rightarrow e\mu)$+jets &  37  ~,~18  & 2   ~,~ 1\\
Misidentified leptons                         &  164 ~,~65  & 116 ~,~ 55\\
\hline
Total background & 1370 ~,~ 120 & 340 ~,~ 88 \\
\hline
\end{tabular}
\caption{\label{t:evtsum}Observed numbers of opposite-sign $e\mu$ events
with one and two $b$-tagged jets ($N_1$ and $N_2$), together with the
estimates of non-$t\bar{t}$ backgrounds and associated systematic uncertainties. 
Uncertainties quoted as 0 are $< 0.5$.}
\end{table}

The distribution of the number of $b$-tagged jets in opposite-sign $e\mu$ 
events is shown in Figure~\ref{f:btags}, and compared to the baseline
and alternative \ttbar\ and background simulation samples. 
The  \ttbar\ contribution is normalised to the theoretical \ttbar\ cross-section prediction at \sxyt\ of 832\,pb.
The agreement between data and simulation in the one and two $b$-tagged bins
used for the cross-section measurement is good. However, the data has
about 40\% more events with three or more $b$-tags than the baseline 
simulation, indicating a mismodelling of events with 
\ttbar\ produced in association with additional heavy-flavour jets, as discussed
further in Section~\ref{sec:syst}. 
There is also an approximately 11\% excess of data over
simulation for events with zero $b$-tagged jets which does not affect the 
measurement, and is compatible with the expected uncertainties in modelling
 $WW$~\cite{ww8TeV} and $Z$+jets production.
Distributions of the number of jets, the jet $\pt$, 
and the electron and muon $|\eta|$ and $\pt$
are shown for opposite-sign $e\mu$ events with at least one $b$-tagged jet in
Figure~\ref{f:dmc}, where the simulation is normalised to the same number of
events as the data. 
In general, the data and simulation agree well.

\begin{figure}[h]
\centering
\includegraphics[width=90mm]{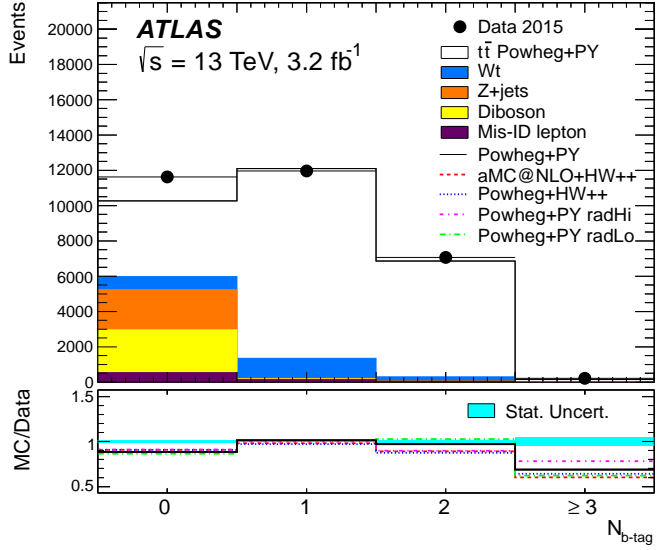}
\caption{\label{f:btags}Distribution of the number of $b$-tagged jets
in preselected opposite-sign $e\mu$ events. The data are shown compared
to the prediction from simulation, broken down into contributions from
\ttbar\ (using the baseline \textsc{Powheg+Pythia6} sample), 
$Wt$ single top, $Z$+jets, dibosons, and events with fake electrons or
muons, normalised to the same integrated luminosity as the data.
The lower part of the figure shows the ratio of simulation to data, using 
various \ttbar\ signal samples, and the shaded band indicates the statistical uncertainty. 
The  \ttbar\ contribution is normalised to the theoretical \ttbar\ cross-section prediction 
at \sxyt\ of 832\,pb.}  
\end{figure}

\begin{figure}[tbp]
  \splitfigure{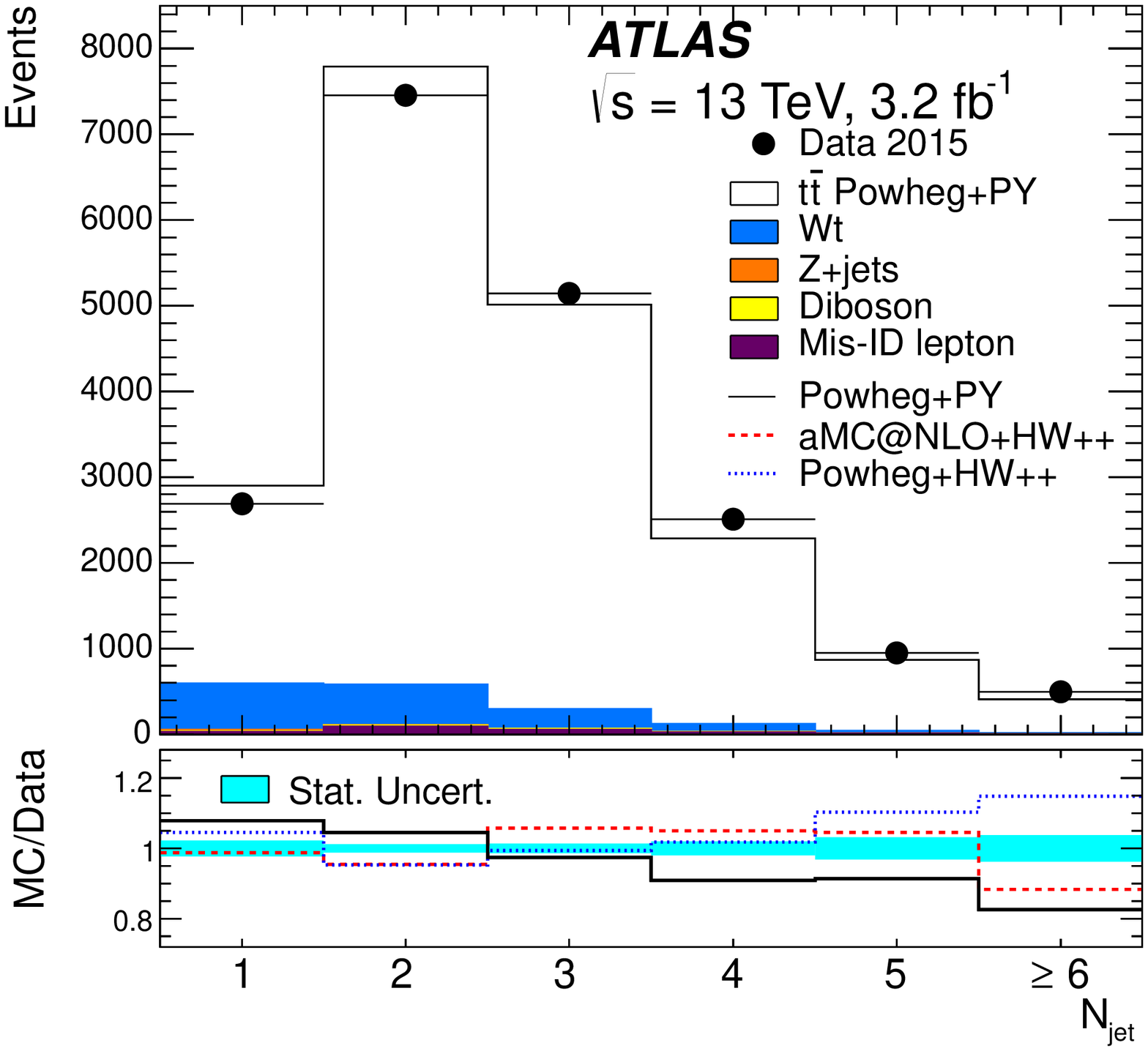}{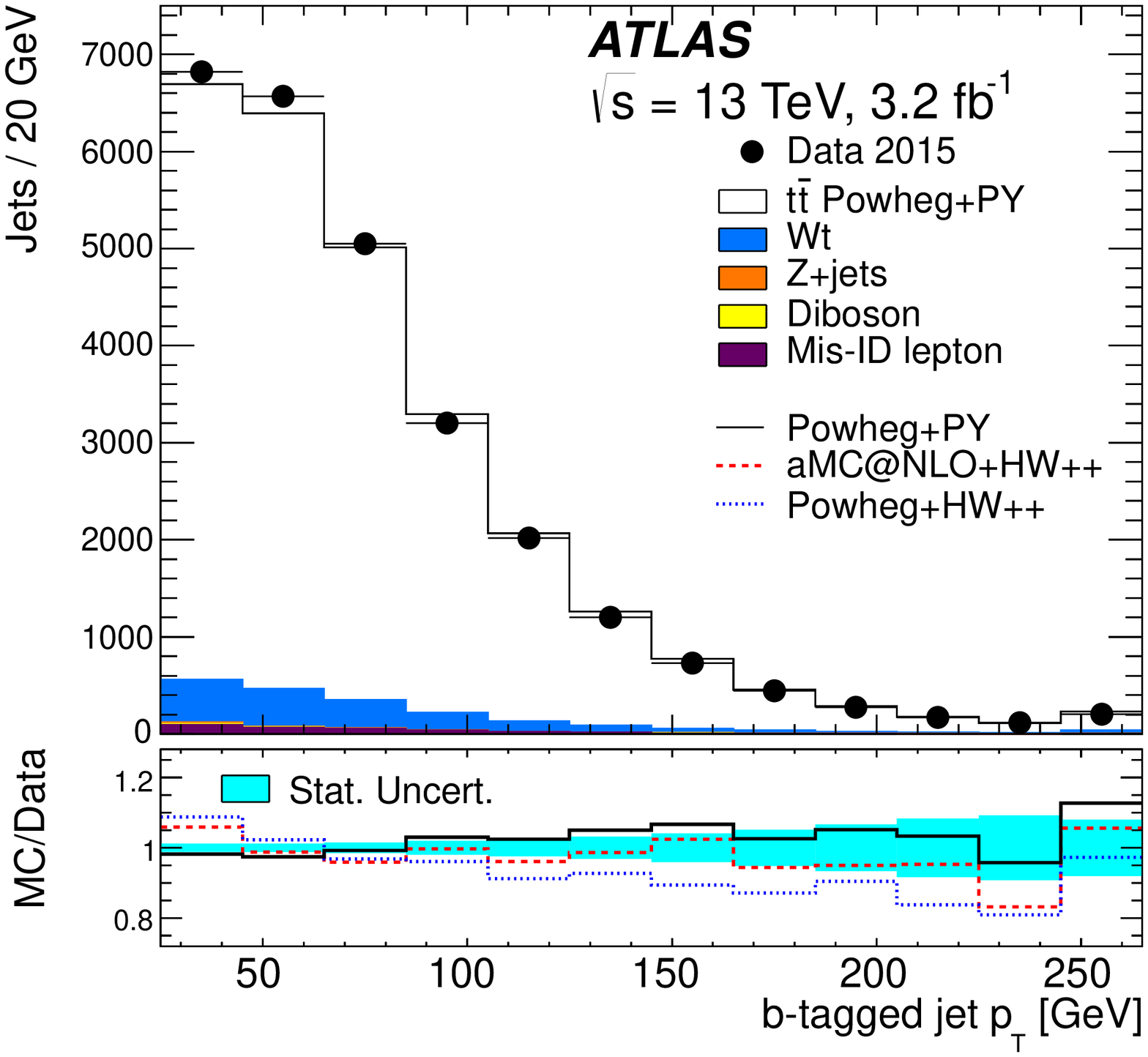}{a}{b}
  \splitfigure{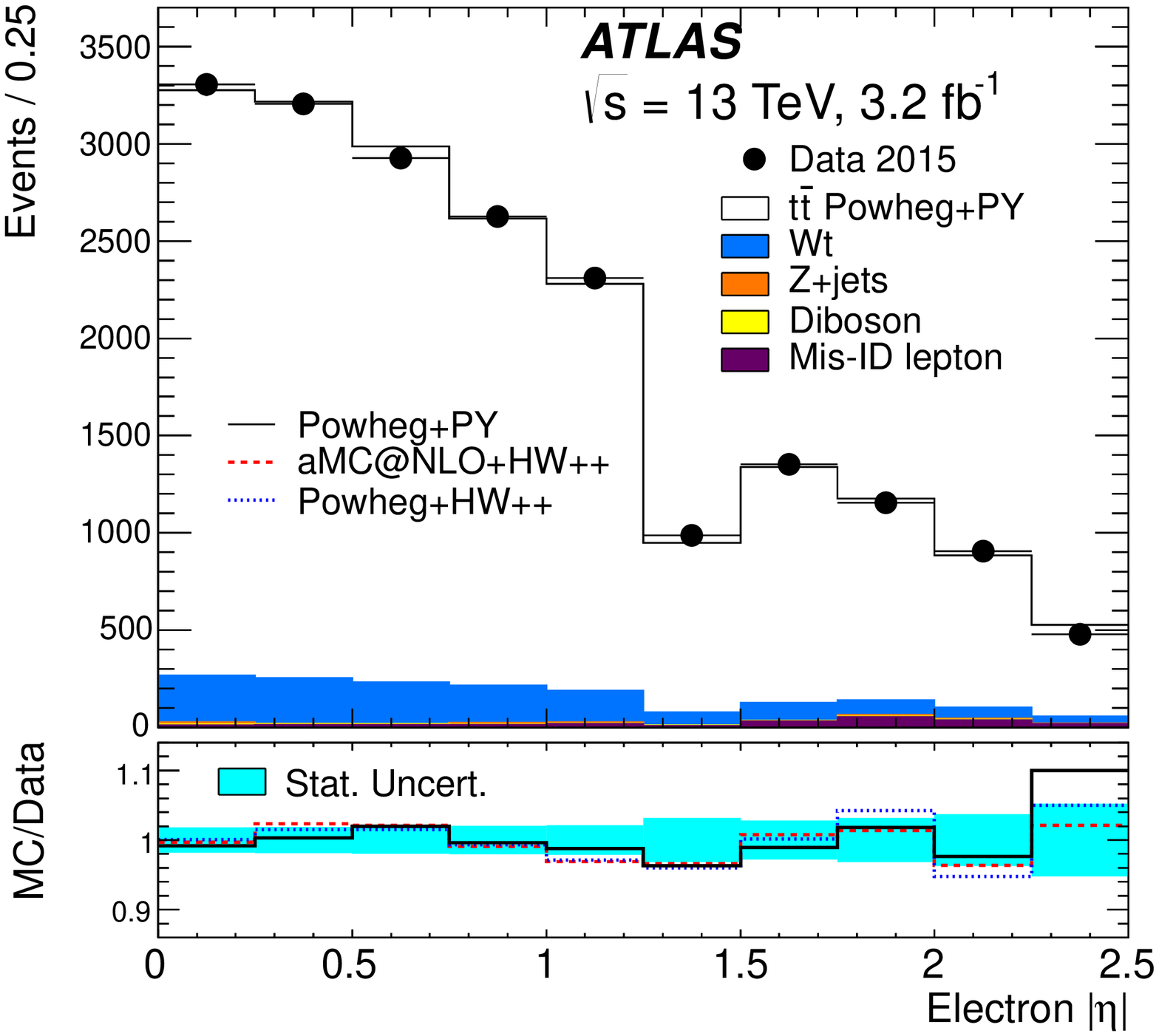}{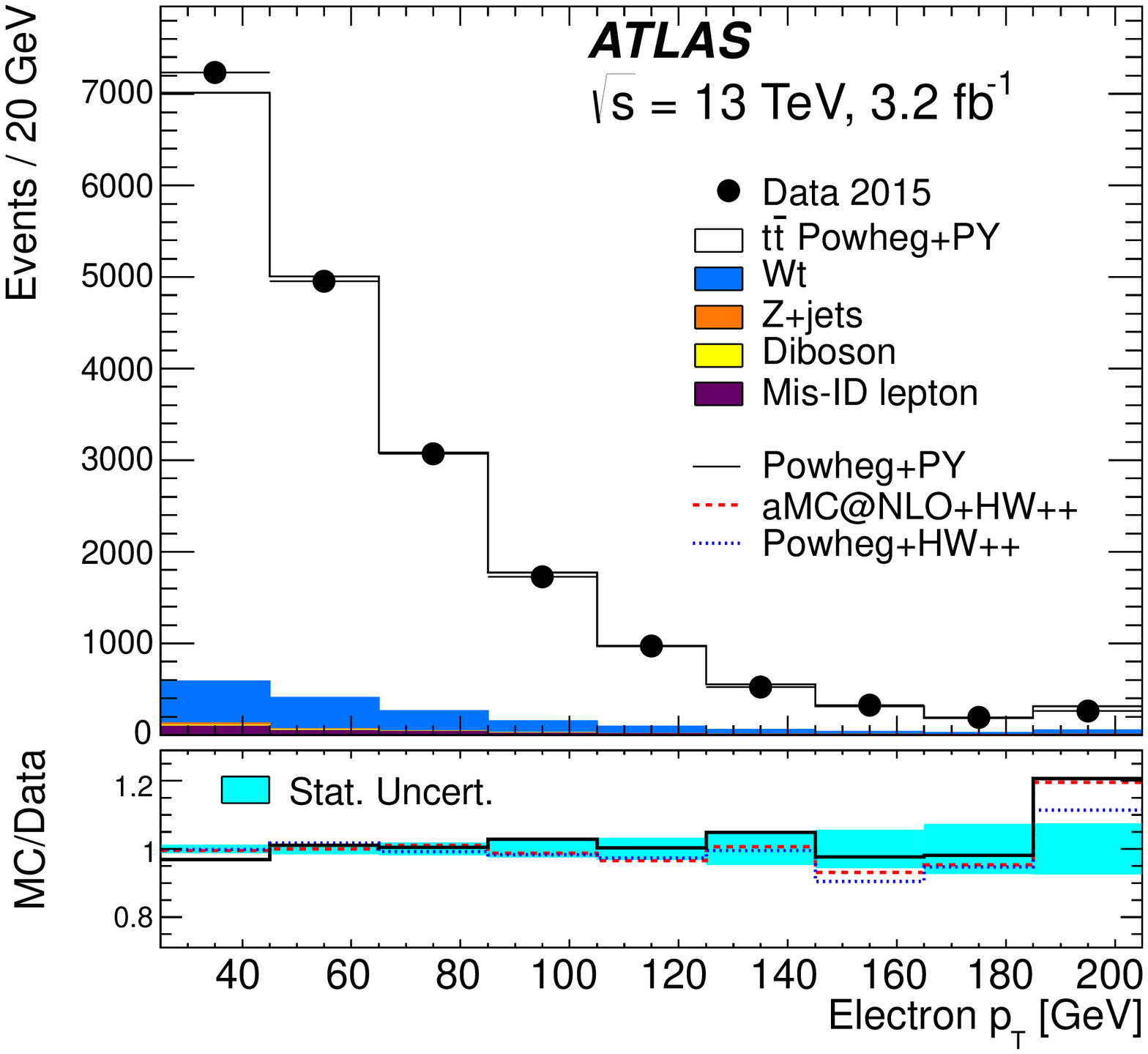}{c}{d}
  \splitfigure{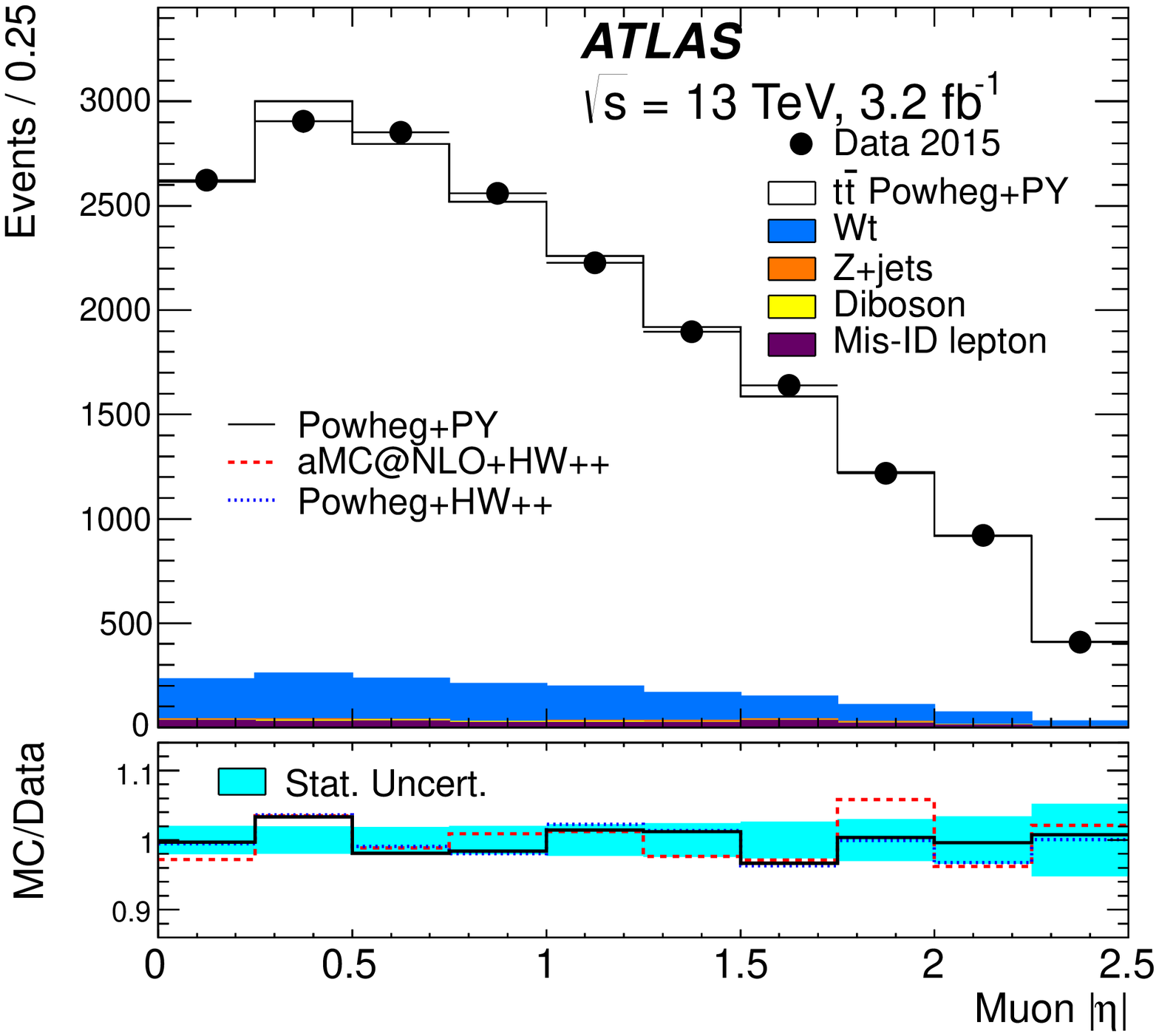}{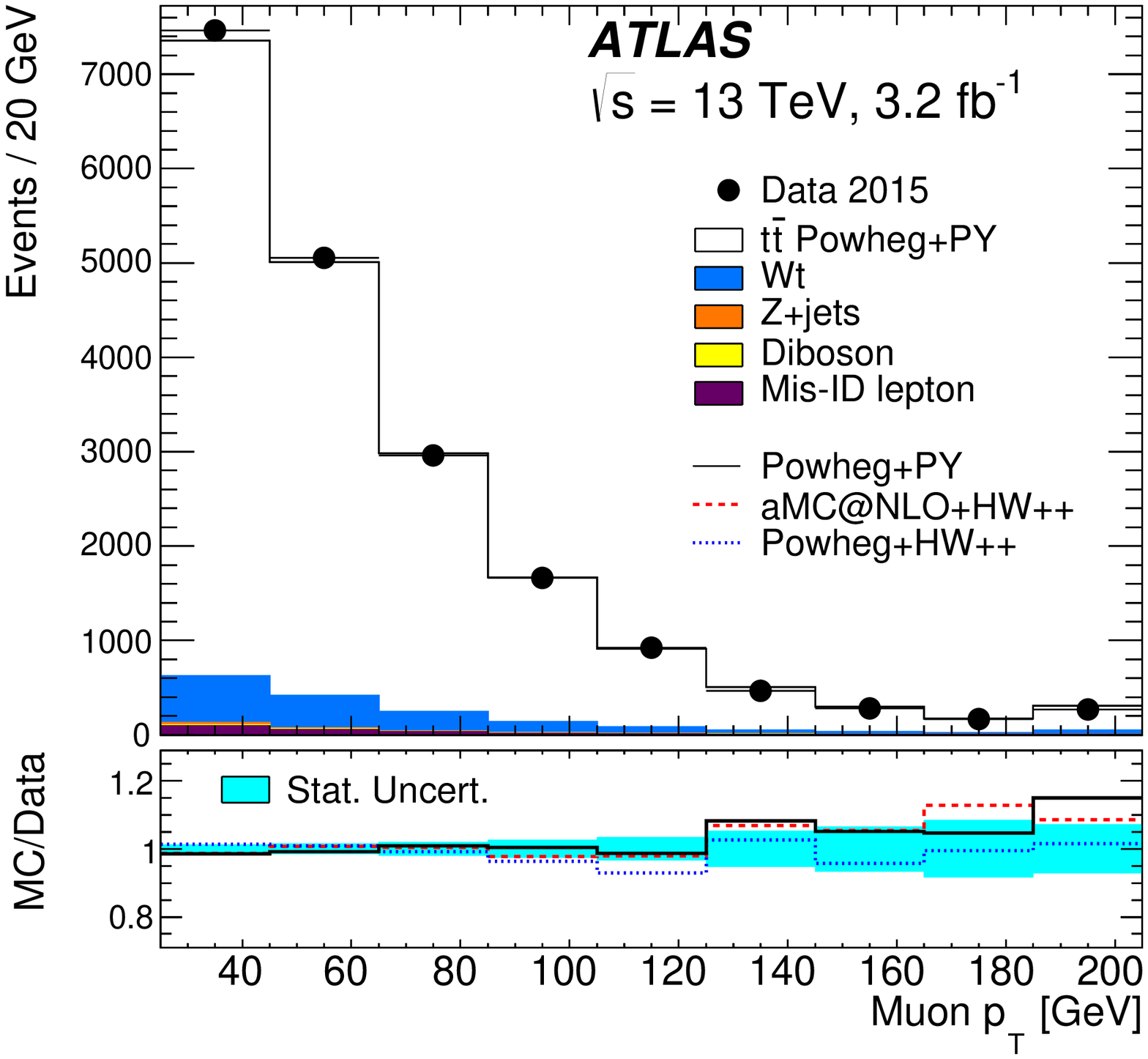}{e}{f}
  \caption{\label{f:dmc}Distributions of 
    (a) the number of jets, 
    (b) the transverse momentum $\pt$ of the $b$-tagged jets,  
    (c) the $|\eta|$ of the electron,
    (d) the $\pt$ of the electron,
    (e) the $|\eta|$ of the muon 
    and 
    (f) the $\pt$ of the muon, 
    in events with an opposite-sign $e\mu$ pair and 
    at least one $b$-tagged jet. The data are compared to the prediction from
    simulation, broken down into contributions from $t\bar{t}$~(using
    the baseline \textsc{Powheg+Pythia6} sample), single top, $Z$+jets,
    dibosons, and events with fake electrons or muons, normalised to the
    same number of entries as the data. 
    The lower parts of the figures show the ratios of simulation to data,
    using various $t\bar{t}$ signal samples, and with the shaded band 
    indicating the statistical uncertainty.
    The last histogram bin includes the overflow.}
\end{figure}

%-------------------------------------------------------------------------------
\section{Background estimation}
\label{sec:back}
%-------------------------------------------------------------------------------
Most background contributions are estimated from simulation.
The $Wt$ single-top background is normalised to the 
approximate NNLO cross-section of $71.7\pm 3.8$\,pb, determined as in 
Ref. \cite{Wttheoxsec}. 
The diboson background normalisation is estimated using
\textsc{Sherpa} as discussed in Section~\ref{sec:dmc}.
The normalisation of the $Z$+jets background, originating from events with a 
$Z\rightarrow\tau\tau\rightarrow e\mu$ decay accompanied by one or two $b$-tagged 
jets, is determined by scaling the \textsc{Sherpa} simulation with scale factors obtained
in  $Z\rightarrow ee$ and $Z\rightarrow\mu\mu$ events as described in Section~\ref{sec:syst}.
% text for fake leptons

The background from events with one real and one misidentified lepton is 
estimated from a combination of data and simulation, using the method employed 
in Ref. \cite{atopxsec}.
Simulation studies show that the samples with 
a same-sign $e\mu$ pair and one or two $b$-tagged jets are dominated by events
with a misidentified lepton, with rates comparable to those in the 
opposite-sign sample. 
The contributions of events with misidentified leptons
are therefore estimated using the same-sign event counts in data
after subtraction of the estimated prompt same-sign contributions, multiplied
by the opposite- to same-sign fake-lepton ratios $R_j$ for $j=1$ and 2  
$b$-tagged jets predicted from simulation. The results are shown in Table~\ref{t:evtsum} and the procedure is illustrated
in Table~\ref{t:fakelept}, which shows the expected breakdown of same-sign 
event counts in terms of prompt-lepton and misidentified-lepton events, and the 
corresponding predictions for misidentified leptons in the opposite-sign
sample with all contributions estimated from simulation. The misidentified-lepton contributions are classified into those
where the electron is from a photon conversion, from the decay of a 
heavy-flavour hadron or from other sources ({\em e.g.}\ a misidentified 
hadron within a jet), or the muon is from a heavy-flavour
decay or other sources ({\em e.g.} a pion or kaon decay).
The values of $R_j$ are taken to be $R_1=1.55\pm 0.50$ 
and $R_2=1.99\pm 0.82$, where the central values are taken from ratios of the
total numbers of misidentified-lepton events in opposite- and same-sign
samples. The uncertainties encompass the different
values of $R_j$ predicted for the various sub-components of the misidentified-lepton 
background separately, allowing the background composition to be significantly different from that predicted by simulation,
where it is dominated by electrons from photon conversions,
followed by electrons and muons from the decays of heavy-flavour hadrons.
A 50\% uncertainty is assigned to the prompt same-sign contribution, which
includes events where the charge of the electron was misidentified 
(denoted by wrong-sign prompt in Table~\ref{t:fakelept}) or right-sign with
two genuine same-sign leptons ({\em e.g.} from $\ttbar W/Z$ production).
The largest uncertainties in the misidentified-lepton background come from
the uncertainties in $R_j$.

% table for fake leptons
% TAGFBL updated 12/2/16 from confplots.fakebg().table()

\begin{table}[tp]
\centering 
\begin{tabular}{l| D{,}{~\pm~}{-1} D{,}{~\pm~}{-1} D{,}{~\pm~}{-1} D{,}{~\pm~}{-1} }\hline
Component &  \multicolumn{1}{c}{OS $1b$} &  \multicolumn{1}{c}{SS $1b$} &  \multicolumn{1}{c}{OS $2b$} &  \multicolumn{1}{c}{SS $2b$} \\
\hline
%Conversion $e$      &  115 ,   5    &   84 ,   5   & 61 ,   3   & 33.7,  1.7  \\
%Heavy-flavour $e$   &  11.3,  1.8   &   9.4,  1.0  & 1.1,  0.3  &  0.9,  0.3  \\
%Other $e$           &   15 ,  13    &   0.6,  0.2  & 3.4,  1.9  &  0.2,  0.1  \\
%Heavy-flavour $\mu$ &   9.5,  0.9   &   5.3,  0.7  & 2.1,  0.5  &  0.5,  0.2  \\
%Other $\mu$         &   3.4,  0.5   &   0.3,  0.2  & 2.8,  0.5  &  0.0,  0.0  \\
%\hline
%Total misidentified & 154,14 &  99,6 & 70,4 &    35,2 \\\hline
%Wrong-sign prompt   &  \multicolumn{1}{c}{~~~~~-} &  30.5,1.6 &  \multicolumn{1}{c}{~~~~~-} & 16.2,1.2 \\
%Right-sign prompt   &  \multicolumn{1}{c}{~~~~~-} &  12.0,0.5 &  \multicolumn{1}{c}{~~~~~-} & 4.4,0.2 \\
%\hline
%Total               &  \multicolumn{1}{c}{~~~~~-} &   142,6 &  \multicolumn{1}{c}{~~~~~-} & 56,2 \\\hline
%Data                &  \multicolumn{1}{c}{~~~~~-} &  \multicolumn{1}{c}{~~~~149} &  \multicolumn{1}{c}{~~~~~-} &  \multicolumn{1}{c}{~~~~79} \\
\hline
Conversion $e$          &   113 ,   5    &    83 ,   5   &    60 ,  3    &   33.3,  1.7  \\
Heavy-flavour $e$       &   11.0,  1.8   &    9.8,  0.9  &    1.1,  0.3  &    0.9,  0.3  \\
Other $e$               &    15 ,   13   &    0.4,  0.2  &    3.3,  1.9  &    0.2,  0.1  \\
Heavy-flavour $\mu$     &    9.5,  0.9   &    5.6,  0.7  &    1.9,  0.4  &    0.5,  0.2  \\
Other $\mu$             &    3.4,  0.5   &    0.3,  0.2  &    2.7,  0.5  &    0.0,  0.0  \\
\hline
Total misidentified & 151,14        &  99,5 & 69,4 &    35,2 \\\hline
Wrong-sign prompt   &  \multicolumn{1}{c}{~~~~~-} &  30.0,1.6 &  \multicolumn{1}{c}{~~~~~-} & 16.0,1.1 \\
Right-sign prompt   &  \multicolumn{1}{c}{~~~~~-} &  11.8,0.5 &  \multicolumn{1}{c}{~~~~~-} & 4.4,0.2 \\
\hline
Total               &  \multicolumn{1}{c}{~~~~~-} &   141,6 &  \multicolumn{1}{c}{~~~~~-} & 55,2 \\\hline
Data                &  \multicolumn{1}{c}{~~~~~-} &  \multicolumn{1}{c}{~~~~149} &  \multicolumn{1}{c}{~~~~~-} &  \multicolumn{1}{c}{~~~~79} \\
\hline
\end{tabular}
\caption{\label{t:fakelept}The expected numbers of events with at least one
misidentified lepton in the one- and two-$b$-tag opposite- and same-sign $e\mu$
samples, broken down into different categories as described in the text. 
For the same-sign samples, the contributions from wrong-sign (where the 
electron charge sign is misreconstructed) and right-sign prompt lepton
events are also shown, and the total expected numbers of events are compared to the data.
The uncertainties are due to simulation statistics, and numbers
quoted as `0.0' are smaller than 0.05.}
\end{table}

The modelling in simulation of the different components of the 
misidentified-lepton background is checked by studying kinematic distributions
of same-sign events, as illustrated for the \pt\ and $|\eta|$ distributions of
the leptons in events with at least one $b$-tagged jet in 
Figure~\ref{f:sslept}. The simulation models the shapes of the distributions
well, but underestimates the number of data events with two $b$-tagged
jets by about 40\%, as shown in Table~\ref{t:fakelept}. This deficit in simulation is 
attributed to a larger rate of misidentified-lepton events in data, which increases the 
estimate of misidentified leptons in the opposite-sign two-$b$-tag sample accordingly. 
The modelling 
is also checked in same-sign control samples with relaxed isolation cuts,
enhancing the contributions of heavy-flavour decays, and similar levels
of agreement were found, giving confidence that the simulation adequately
models the different sources of misidentified leptons in the selected sample.

% TAGFIG updated 12/2/16 RH, confplots.ttproj().conffig3([1-4])
\begin{figure}[tbp]
\splitfigure{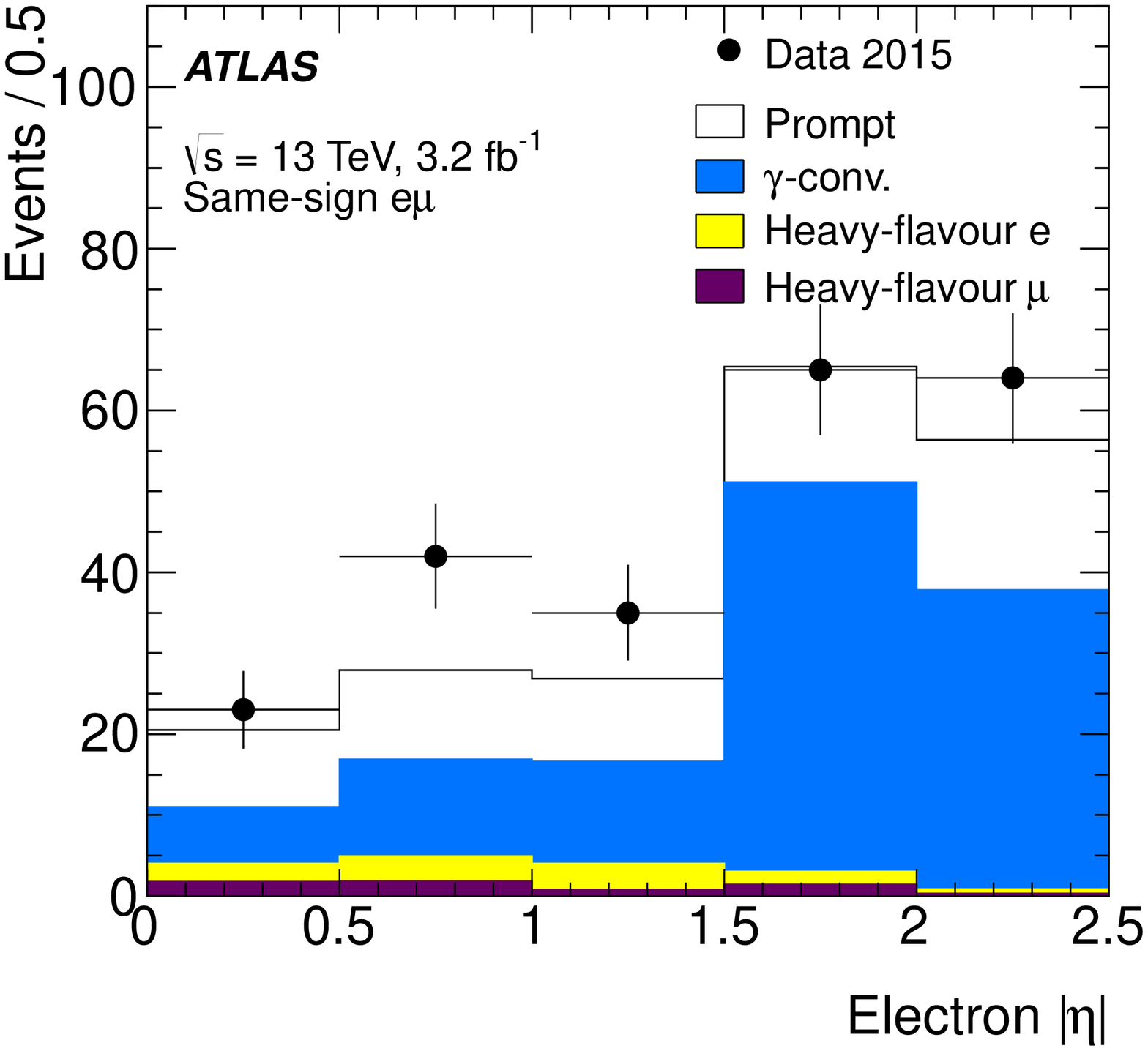}{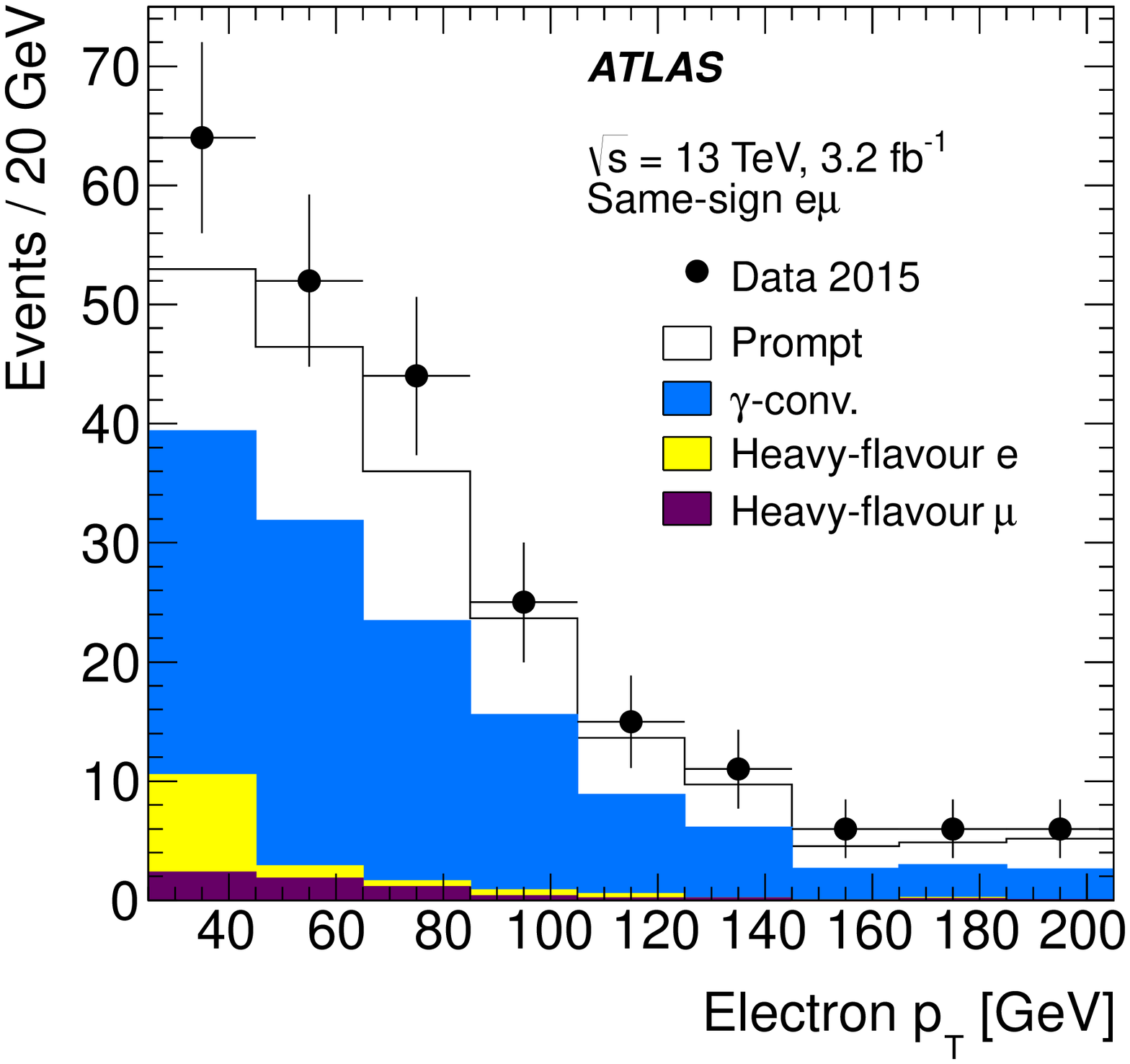}{a}{b}
\splitfigure{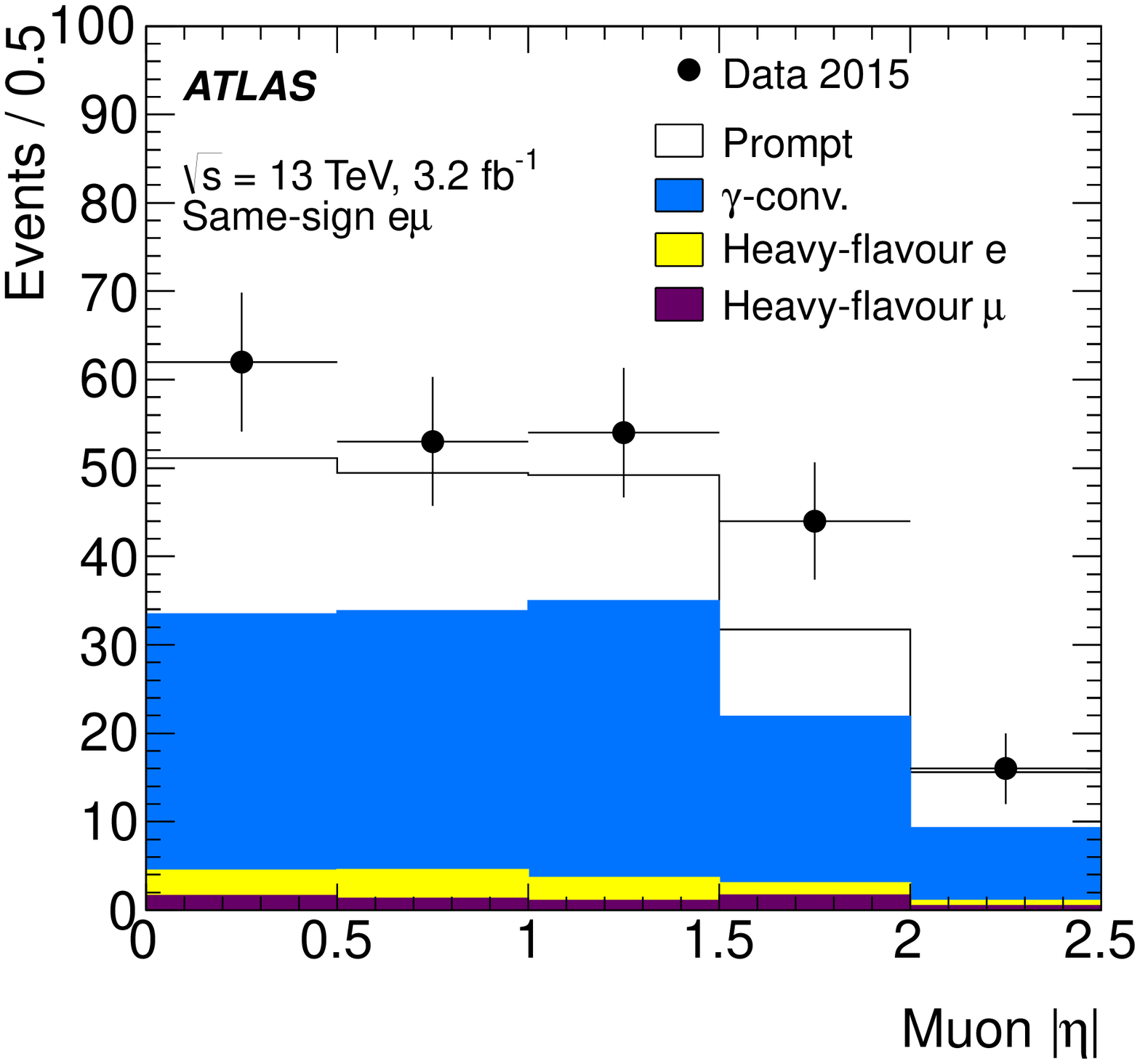}{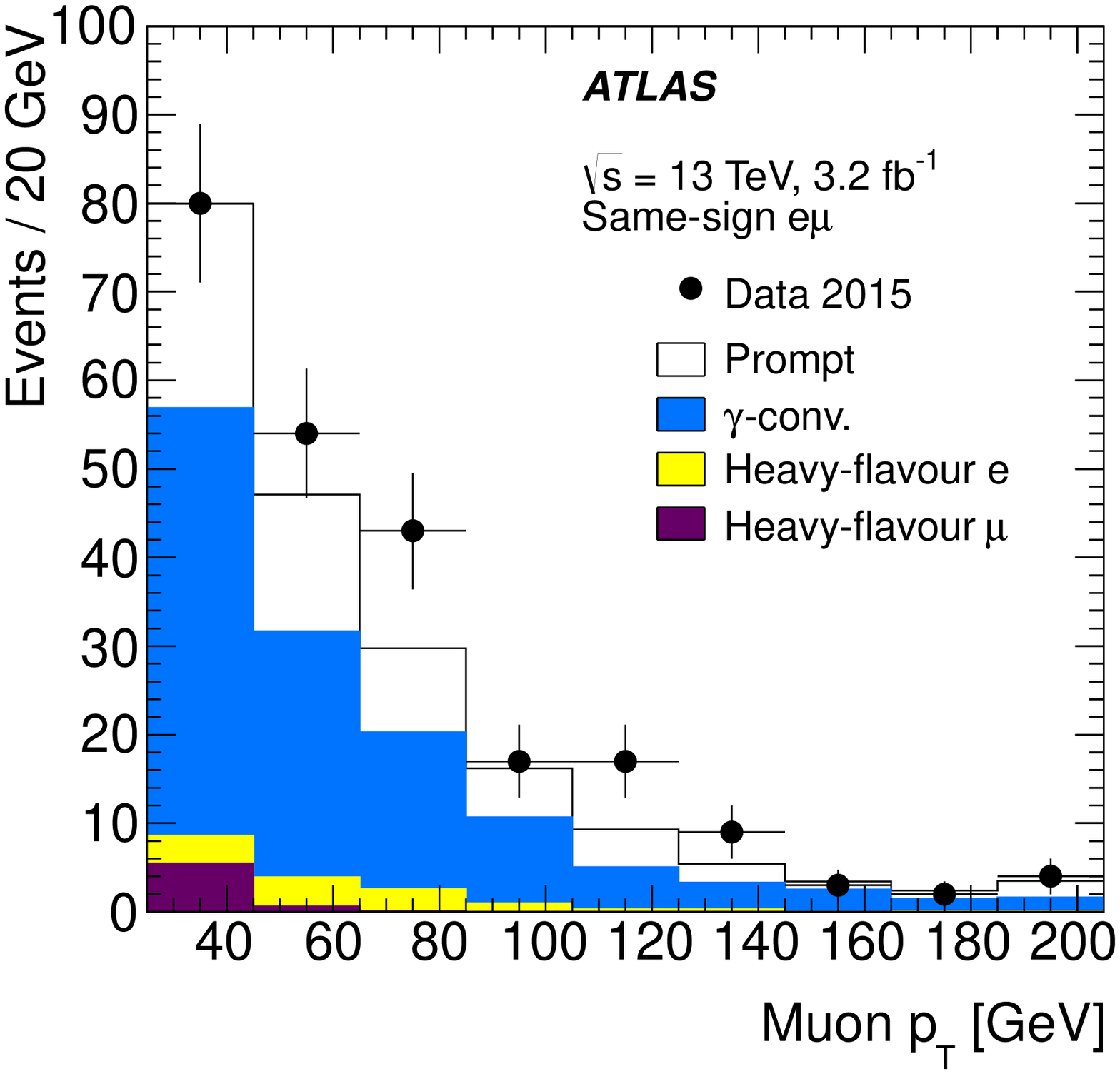}{c}{d}
\caption{\label{f:sslept}Distributions of electron and muon $|\eta|$ and
\pt\ in same-sign $e\mu$ events with at least one $b$-tagged jet. The
simulation prediction is normalised to the same integrated luminosity as the
data, and broken down into contributions where both leptons are prompt, 
or one is a misidentified lepton from a photon conversion or heavy-flavour
decay. In the \pt\ distributions, the last bin includes the overflow.}
\end{figure}

%%%%%%%%%%%%%%

%-------------------------------------------------------------------------------
\section{Systematic uncertainties}
\label{sec:syst}
%-------------------------------------------------------------------------------

The systematic uncertainties in the extracted cross-sections, \xtt\ and $\sigma_{t\bar{t}}^{\textrm fid}$, are 
shown in Table~\ref{t:systxtt}, together with their effects (where relevant)
on the 
\ttbar\ preselection efficiency \epsem, tagging correlation \cb\ and reconstruction efficiency \gem. Each
source of uncertainty is evaluated by repeating the cross-section extraction with all 
relevant input parameters simultaneously changed by $\pm 1$ standard deviation.
Correlations between input parameters
(in particular significant anti-correlations between \epsem\ and \cb\ which 
contribute with opposite signs to \xtt) are thus taken into account. The
total uncertainties are calculated by adding the effects
of all the individual systematic components in quadrature, assuming them
to be independent. The sources of systematic uncertainty are discussed
in detail below.

\begin{table}[tp]
\centering
\begin{tabular}{l|rrr}\noalign{\smallskip}\hline\noalign{\smallskip}

Uncertainty  (inclusive \xtt) & $\Delta\epsem/\epsem$ [\%] & $\Delta\cb/\cb$ [\%]& $\Delta\xtt/\xtt$ [\%] \\
\noalign{\smallskip}\hline\noalign{\smallskip}
Data statistics & & &   0.9 \\
\noalign{\smallskip}\hline\noalign{\smallskip}
\ttbar\ NLO modelling &   0.7 &  $-0.1$ &   0.8 \\
\ttbar\ hadronisation &  $-2.4$ &   0.4 &   2.8 \\
Initial- and final-state radiation &  $-0.3$ &   0.1 &   0.4 \\
\ttbar\ heavy-flavour production & -~~ &   0.4 &   0.4 \\
Parton distribution functions &   0.5 & -~~ &   0.5 \\
Single-top modelling & -~~ & -~~ &   0.3 \\
Single-top/\ttbar\ interference & -~~ & -~~ &   0.6 \\
Single-top $Wt$ cross-section & -~~ & -~~ &   0.5 \\
Diboson modelling & -~~ & -~~ &   0.1 \\
Diboson cross-sections & -~~ & -~~ &   0.0 \\
$Z$+jets extrapolation & -~~ & -~~ &   0.2 \\
Electron energy scale/resolution &   0.2 &   0.0 &   0.2 \\
Electron identification &   0.3 &   0.0 &   0.3 \\
Electron isolation &   0.4 & -~~ &   0.4 \\
Muon momentum scale/resolution &  $-0.0$ &   0.0 &   0.0 \\
Muon identification &   0.4 &   0.0 &   0.4 \\
Muon isolation &   0.2 & -~~ &   0.3 \\
Lepton trigger &   0.1 &   0.0 &   0.2 \\
Jet energy scale &   0.3 &   0.1 &   0.3 \\
Jet energy resolution &  $-0.1$ &   0.0 &   0.2 \\
$b$-tagging & -~~ &   0.1 &   0.3 \\
Misidentified leptons & -~~ & -~~ &   0.6 \\
\noalign{\smallskip}\hline\noalign{\smallskip}
Analysis systematics &   2.7 &   0.6 &   3.3 \\
\noalign{\smallskip}\hline\noalign{\smallskip}
Integrated luminosity & -~~ & -~~ &   2.3 \\
LHC beam energy & -~~ & -~~ &   1.5 \\
\noalign{\smallskip}\hline\noalign{\smallskip}
Total uncertainty &   2.7 &   0.6 &   4.4 \\
\noalign{\smallskip}\hline\noalign{\smallskip}
\noalign{\smallskip}\hline\noalign{\smallskip}
Uncertainty (fiducial \xfid) &  $\Delta\gem/\gem$ [\%]& $\Delta\cb/\cb$ [\%]&$\Delta\xfid/\xfid$ [\%]\\
\noalign{\smallskip}\hline\noalign{\smallskip}
\ttbar\ NLO modelling &   0.5 &  $-0.1$ &   0.6  \\
\ttbar\ hadronisation &  $-1.6$ &   0.4 &   1.9  \\
Parton distribution functions &   0.1 & -~~ &   0.1  \\
Other uncertainties (as above) &  0.8 &  0.4 &   1.5\\
\noalign{\smallskip}\hline\noalign{\smallskip}
Analysis systematics (\xfid) &   1.8 &   0.6 &   2.5 \\
\noalign{\smallskip}\hline\noalign{\smallskip}
Total uncertainty (\xfid) &   1.8 &   0.6 &   3.9 \\
\noalign{\smallskip}\hline\noalign{\smallskip}

\end{tabular}
\caption{\label{t:systxtt}Summary of the systematic uncertainties
in \epsem, \gem\ and \cb\ (with their relative signs where relevant), and the statistical, systematic, luminosity and  beam energy
uncertainties in the measured $t\bar{t}$ cross-section $\sigma_{t\bar{t}}$  at $\sqrt{s}=13$~TeV. 
All uncertainties from the inclusive cross-section measurement apply to the fiducial measurement; in 
the lower part of the table only the systematic uncertainties
that are different for the measurement of the fiducial cross-section $\sigma_{t\bar{t}}^{\textrm fid}$ are given,
together with the total analysis systematic uncertainties and total uncertainties in $\sigma_{t\bar{t}}^{\textrm fid}$.
Uncertainties quoted as `0.0' are smaller than 0.05\%, whilst `-' indicates that the
corresponding uncertainty is not applicable.
}
\end{table}

\begin{description}

\item[\ttbar\ modelling:] 
The modelling uncertainties in \epsem\ and \cb\ due to the choice of \ttbar\ generator 
are assessed by comparing the predictions of the baseline 
\textsc{Powheg\,+\,Pythia6} sample with the various alternative samples discussed 
in Section~\ref{sec:dmc}. %, all processed using fast simulation. 
Three separate uncertainties are considered: the NLO generator uncertainty (evaluated by
considering the relative difference between \textsc{MadGraph5\_aMC@NLO\,+\,Herwig++} and \textsc{Powheg\,+\,Herwig++}),
the parton shower and hadronisation uncertainty (evaluated by considering the relative difference between \textsc{Powheg\,+\,Pythia6}
and \textsc{Powheg\,+ \,Herwig++}), and the radiation uncertainty (evaluated by considering half the relative difference
between the \textsc{Powheg\,+\,Pythia6} samples with more or less radiation).
The prediction for \epsem\ is found to be particularly sensitive to the amount of hadronic
activity near the leptons, which strongly affects the efficiency of the lepton isolation 
requirements described in Section~\ref{sec:objev}. These isolation efficiencies are therefore measured directly from
data, as discussed below, and thus no modelling uncertainty is considered for the lepton isolation. Motivated by the level of agreement for events with at least three
$b$-tags seen in Figure~\ref{f:btags}, an additional uncertainty in \cb\
is determined by calculating  in data and simulation the ratio \rtt\ of the number of events with at 
least three $b$-tagged jets to the number with at least two. The baseline
simulation sample is reweighted to change the fraction of events with
at least three $b$-jets at generator level, which effectively changes the
\ttbar\ plus heavy-flavour fraction and the values of both \cb\ and \rtt.
A linear relation between changes in \cb\ and \rtt\ is found, and used
to translate the difference between the \rtt\ values found in data ($3.1\pm 0.2$\%) and simulation ($2.21\pm 0.05$\%) 
to a shift in \cb\ of 0.39\%.
This shift is treated as an additional uncertainty in \cb\ due to the
modelling of heavy-flavour production in \ttbar\ events, uncorrelated to the
NLO, hadronisation and radiation uncertainties discussed above.

\item[Parton distribution functions:]  
The uncertainties in \epsem\ and \cb\ due to limited knowledge of the proton PDFs are evaluated by reweighting
simulated events produced with \textsc{MadGraph5\_aMC@NLO} using the error sets of the NNPDF~3.0 PDF sets~\cite{nnpdfffn3.0}. 
The eigenvectors consist of a central PDF and 100 Monte Carlo replicas, 
for which the root mean square was taken to calculate the uncertainty.
The \textsc{MadGraph5\_aMC@NLO} sample was produced with CT10; therefore the cross-section
was corrected for the relative difference between the central prediction of CT10 and NNPDF~3.0, which is about~1\%.
The uncertainty using the PDF4LHC Run-2 recommendations with 100 eigenvectors~\cite{pdflhcrun2} is very similar to that obtained with NNPDF~3.0.

\item[Single-top modelling:] The uncertainties related to $Wt$ single-top 
modelling are assessed by comparing the predictions of \textsc{Powheg\,+\,Pythia6}
and \textsc{Powheg\,+\,Herwig++} and  considering the relative difference,  comparing the diagram removal and diagram subtraction schemes for dealing with
the interference between the \ttbar\ and $Wt$ final states, and also considering half the relative difference
between the \textsc{Powheg\,+\,Pythia6} samples with more or less radiation.
Production of single top quarks via the $t$- and $s$-channels gives rise to 
final states with only one prompt lepton, and is accounted for as part of the 
misidentified-lepton background.

\item[Diboson modelling:] 
The uncertainties in the background contributions from
dibosons with one or two additional $b$-tagged jets were assessed by comparing
the baseline prediction from \textsc{Sherpa} with that of \textsc{Powheg+Pythia8}. 
These uncertainties have a limited effect on the cross-section
measurement due to the small number of diboson background events.

\item[Background cross-sections:] The uncertainties in the $Wt$ single-top 
and diboson cross-sections are taken to be 5.3\%~\cite{Wttheoxsec} and
6\%~\cite{Dibxsec}, based on the corresponding theoretical predictions.

\item[$Z$+jets extrapolation:] 
The cross-sections for $Z$+jets and especially $Z$+heavy-flavour jets are subject to large 
theoretical uncertainties, making purely simulation-based estimates
unreliable. This background was therefore determined by measuring the
rates of $Z\rightarrow ee$ and $Z\rightarrow\mu\mu$ events with one and 
two $b$-tagged jets in both data and simulation, and using the resulting ratio 
to scale the simulation estimate of background from $Z\rightarrow\tau\tau$+jets. 
The $Z$+jets background prediction from simulation
was scaled by 1.1 for the background with one $b$-tagged jet
%the one $b$-tagged jet background 
and by 1.2 for the background with two $b$-tagged jets.
%the two $b$-tagged jets background. 
A 50\% uncertainty was applied to the $Z$+jets contributions which cover the differences observed on 
the event yields comparing Z+jets \textsc{Sherpa} vs \textsc{Powheg+Pythia8}.

\item[Lepton-related uncertainties:] The modelling of the 
electron and muon trigger efficiencies, identification efficiencies, energy scales and resolutions
are studied using $Z\rightarrow ee$ and $Z\rightarrow\mu\mu$ decays in 
data and simulation.
Small corrections are applied to the simulation to improve  the agreement with the response observed in data. These corrections have associated 
uncertainties that are propagated to the cross-section measurement. 
The uncertainty in the trigger efficiency is small compared to those
for electron or muon identification since most events are triggered 
redundantly by both leptons.
The efficiency of the lepton isolation requirements was measured directly in data 
\ttbar\ events, thus including the effects of pile-up, by relaxing the cuts alternately on electrons and muons as in Ref.~\cite{atopxsec}.
The results, after the correction for the contamination from misidentified leptons estimated using the
same-sign $e\mu$ samples as described in Section~\ref{sec:back}, showed that the
baseline \textsc{Powheg+Pythia6} simulation overestimates the efficiencies of the isolation requirements 
by about 0.2\% for both the electrons and muons. These corrections were applied to  \epsem\, and the corresponding 
uncertainties are dominated by the subtraction of misidentified leptons.

\item[Jet-related uncertainties:] Although the efficiency to reconstruct and
$b$-tag jets from \ttbar\ events is extracted from the data, uncertainties
in the jet energy scale, energy 
resolution and reconstruction efficiency affect the
backgrounds estimated from simulation and the estimate of the tagging 
correlation \cb. They also have a small effect on \epsem\ via the 
lepton--jet $\Delta R$ separation cuts. 
 The jet energy scale is varied in simulation 
according to the uncertainties derived from the \sxvt\ simulation and data calibration, extrapolated to \sxyt~\cite{jesrun2}. 
The uncertainties are evaluated using a model with 19 separate orthogonal components and the resulting variations were added in quadrature. %which were then added in quadrature. 
The jet energy resolution uncertainty is also 
assessed using \sxvt\ data, and extrapolated to \sxyt. 

\item[$b$-tagging uncertainties:] 
The correlation factor \cb\  depends weakly
on the $b$-tagging and mistagging efficiencies predicted by the simulation,
as it is evaluated from the numbers of events with
one and two $b$-tagged jets. 
The uncertainties are determined from \sxvt\ data, with additional uncertainties to account for the presence of the newly-installed insertable B-layer detector (IBL)~\cite{ibl}
and the extrapolation to \sxyt.
Since the definition of \epsem\ does not involve $b$-tagged jets, it has no
$b$-tagging or mistagging-related uncertainties.

\item[Misidentified leptons:] 
The uncertainties in the number of events with misidentified leptons
in the one and two $b$-tagged samples are derived from the
statistical uncertainties in the numbers of same-sign lepton events, 
the systematic uncertainties in the opposite- to same-sign ratios $R_j$,
and the uncertainties in the numbers of prompt same-sign events,  as discussed
in detail in Section~\ref{sec:back}.

\item[Integrated luminosity:] 
The uncertainty in the integrated luminosity is \lumirunc. 
It is derived, following a methodology similar to that detailed in Ref.~\cite{DAPR-2011-01}, % and \cite{DAPR-2015-01}, 
from a calibration of the luminosity scale using $x$--$y$ beam-separation 
scans performed in August 2015.
The effect on the cross-section measurement is slightly larger than 
\lumirunc\ because the $Wt$ single-top and diboson
backgrounds are evaluated from simulation, so they are also sensitive to the
assumed integrated luminosity.

\item[LHC beam energy:] The LHC beam energy during the 2012 $pp$  run
was calibrated to be $0.30\pm 0.66$\% smaller than the 
nominal value of 4\,\TeV\ per beam, using the revolution frequency difference
of protons and lead ions during $p$+Pb runs in early 2013~\cite{ebeam}. 
This relative uncertainty is also applicable to the 2015 $pp$ run.
Since this calibration is compatible with the nominal centre-of-mass energy of 13\,TeV,
no correction is applied to the measured \xtt\ value. However, an 
uncertainty of 1.5\%, corresponding to the expected change in \xtt\ for
a 0.66\% change in  centre-of-mass energy, is quoted separately for the final result.

\item[Top quark mass:]
Alternative \ttbar\ samples generated with different $\mtop$ from 170 to 177.5\,GeV are used to quantify 
the dependence of the acceptance for \ttbar\ events on the assumed \mtop\ value. The level of $Wt$ single-top background based on the 
change of the $Wt$ cross-section for the same mass range is also considered. 
The  \ttbar\ acceptance and background effects partially cancel, and the final 
dependence of the result on the assumed \mtop\ value is
determined to be $\dmtop=\dmtopval$. The result of the analysis is reported for a top quark mass 
of 172.5 GeV, and the small dependence of the cross-section 
on the assumed mass is not included in the total systematic uncertainty.

\end{description}

The total systematic uncertainties in \epsem, \cb, \gem\ and the fitted values
of \xtt\ and $\sigma_{t\bar{t}}^{\textrm fid}$ are shown in Table~\ref{t:systxtt}, and the total
systematic uncertainties in the individual background components
are shown in Table~\ref{t:evtsum}.
The dominant uncertainties in the cross-section result come from the luminosity determination
and \ttbar\ modelling, in particular from the \ttbar\ shower and hadronisation uncertainty.

% All figures and tables should appear before the summary and conclusion.
% The package placeins provides the macro \FloatBarrier to achieve this.
\FloatBarrier
%-------------------------------------------------------------------------------
\section{Results and conclusions}
\label{sec:result}
%-------------------------------------------------------------------------------

The inclusive \ttbar\ production cross-section is measured in the dilepton $\ttbar\rightarrow e\mu\nu\nubar\bbbar$ 
decay channel using \intlumi\,\ifb\ of $\sqrt{s}$ = 13 TeV $pp$ collisions recorded by the ATLAS detector at the LHC.
The numbers
of opposite-sign $e\mu$ events with one and two $b$-tagged jets are
counted, allowing a simultaneous determination of the \ttbar\ cross-section
\xtt\ and the probability to reconstruct and $b$-tag a jet from a \ttbar\
decay. Assuming a top quark mass of $\mtop=172.5$\,GeV, the result is:
\begin{center}
\xtt\ = \ttxval\ $\pm$ \ttxstat\ (stat) $\pm$ \ttxsyst\ (syst) $\pm$ \ttxlumi\ (lumi) $\pm$ \ttxebeam\ (beam)~pb,
\end{center}
where the four uncertainties are due to data statistics, experimental and
theoretical systematic effects, the integrated luminosity and the LHC beam energy, 
giving a total relative uncertainty of \ttxrel. %consistent with the CMS 13 TeV result~\cite{CMSxsec13TeV}. 
The combined probability for a jet from a top quark decay to be within the detector 
acceptance and tagged as a $b$-jet
is measured to be $\epsb=0.559 \pm 0.004 \pm 0.003$, where the first error is 
statistical and the second systematic, in fair agreement with the nominal prediction from simulation of 0.549.

This cross-section measurement is consistent with the theoretical prediction based on NNLO+NNLL calculations of 
$832^{+40}_{-46}$\,pb at $\mtop=172.5$\,GeV.
Figure~\ref{f:sqrts} shows the result of this \xtt\ measurement together with the most precise 
ATLAS results at $\sqrt{s}=7$ and 8\,TeV \cite{atopxsec}.
The data are compared to the NNLO+NNLL predictions as a function of the centre-of-mass energy. %$\sqrt{s}$.
The result is also consistent with a recent measurement by CMS at $\sqrt{s}=13$~\TeV\ using a smaller data sample~\cite{CMSxsec13TeV}.
\begin{figure}[tp]
\centering
\includegraphics[width=120mm]{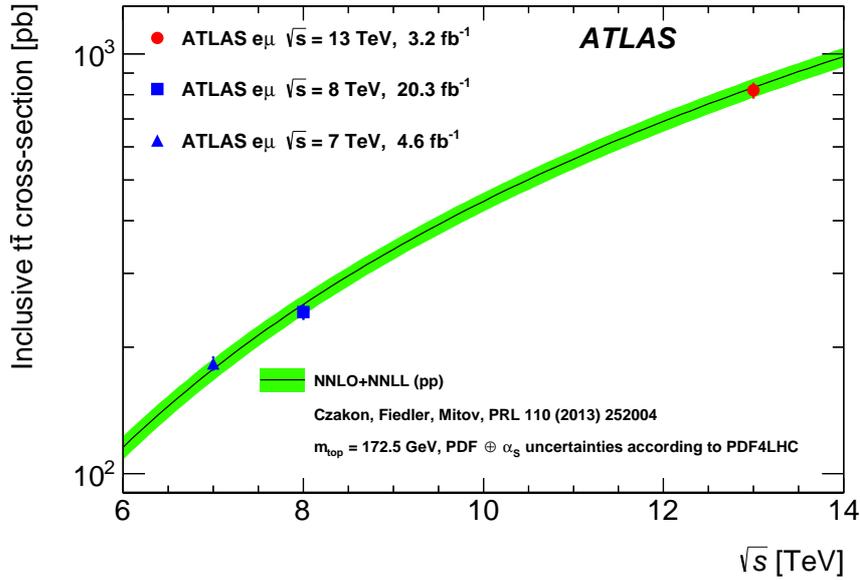}
\caption{\label{f:sqrts} 
Cross-section for \ttbar\ pair production in $pp$ collisions as 
a function of centre-of-mass energy. ATLAS results in the dilepton $e\mu$ 
channel at $\sqrt{s}=13$, 8 and 7\,TeV are compared to the NNLO+NNLL theoretical predictions.
}  
\end{figure}

The measured fiducial cross-section \xfid\ for a \ttbar\ event producing an $e\mu$ pair, each lepton
originating directly from $t\rightarrow W \rightarrow \ell $ or via a leptonic $\tau$ decay   $t\rightarrow W \rightarrow \tau \rightarrow \ell $
and satisfying $p_{\textrm T}>25$\,GeV and $|\eta|<2.5$ is:
\begin{center}
\xfid\ = \xfidval\ $\pm$ \xfidstat\ (stat) $\pm$ \xfidsyst\ (syst) $\pm$ \xfidlumi\ (lumi) $\pm$ \xfidebeam\ (beam)~pb,
\end{center}
with uncertainties due to data statistics, systematic effects, the knowledge of the integrated luminosity and the LHC beam energy, corresponding to a total 
relative uncertainty of \xfidrel\ and an internal systematic uncertainty
excluding the luminosity and the LHC beam energy of~\xfidrelsyst. 
The breakdown of the systematic uncertainties is presented in Table~\ref{t:systxtt}.
Overall, the analysis systematic uncertainties in the fiducial cross-section are smaller than those in the inclusive cross-section, 
due to the substantial reductions in the PDF and hadronisation uncertainties  that contribute significantly to both the acceptance \aem\ and reconstruction efficiency \gem.

%%-------------------------------------------------------------------------------
\section*{Acknowledgements}
%%-------------------------------------------------------------------------------
%
% Acknowledgements for papers with collision data
% Version 2-Aug-2016

% Standard acknowledgements start here
%----------------------------------------------
We thank CERN for the very successful operation of the LHC, as well as the
support staff from our institutions without whom ATLAS could not be
operated efficiently.

We acknowledge the support of ANPCyT, Argentina; YerPhI, Armenia; ARC, Australia; BMWFW and FWF, Austria; ANAS, Azerbaijan; SSTC, Belarus; CNPq and FAPESP, Brazil; NSERC, NRC and CFI, Canada; CERN; CONICYT, Chile; CAS, MOST and NSFC, China; COLCIENCIAS, Colombia; MSMT CR, MPO CR and VSC CR, Czech Republic; DNRF and DNSRC, Denmark; IN2P3-CNRS, CEA-DSM/IRFU, France; GNSF, Georgia; BMBF, HGF, and MPG, Germany; GSRT, Greece; RGC, Hong Kong SAR, China; ISF, I-CORE and Benoziyo Center, Israel; INFN, Italy; MEXT and JSPS, Japan; CNRST, Morocco; FOM and NWO, Netherlands; RCN, Norway; MNiSW and NCN, Poland; FCT, Portugal; MNE/IFA, Romania; MES of Russia and NRC KI, Russian Federation; JINR; MESTD, Serbia; MSSR, Slovakia; ARRS and MIZ\v{S}, Slovenia; DST/NRF, South Africa; MINECO, Spain; SRC and Wallenberg Foundation, Sweden; SERI, SNSF and Cantons of Bern and Geneva, Switzerland; MOST, Taiwan; TAEK, Turkey; STFC, United Kingdom; DOE and NSF, United States of America. In addition, individual groups and members have received support from BCKDF, the Canada Council, CANARIE, CRC, Compute Canada, FQRNT, and the Ontario Innovation Trust, Canada; EPLANET, ERC, FP7, Horizon 2020 and Marie Sk{\l}odowska-Curie Actions, European Union; Investissements d'Avenir Labex and Idex, ANR, R{\'e}gion Auvergne and Fondation Partager le Savoir, France; DFG and AvH Foundation, Germany; Herakleitos, Thales and Aristeia programmes co-financed by EU-ESF and the Greek NSRF; BSF, GIF and Minerva, Israel; BRF, Norway; Generalitat de Catalunya, Generalitat Valenciana, Spain; the Royal Society and Leverhulme Trust, United Kingdom.

The crucial computing support from all WLCG partners is acknowledged gratefully, in particular from CERN, the ATLAS Tier-1 facilities at TRIUMF (Canada), NDGF (Denmark, Norway, Sweden), CC-IN2P3 (France), KIT/GridKA (Germany), INFN-CNAF (Italy), NL-T1 (Netherlands), PIC (Spain), ASGC (Taiwan), RAL (UK) and BNL (USA), the Tier-2 facilities worldwide and large non-WLCG resource providers. Major contributors of computing resources are listed in Ref.~\cite{ATL-GEN-PUB-2016-002}.
%----------------------------------------------

%-------------------------------------------------------------------------------
%\clearpage
%\appendix
%\part*{Appendix}
%\addcontentsline{toc}{part}{Appendix}
%%-------------------------------------------------------------------------------
%
%In a paper, an appendix is used for technical details that would otherwise disturb the flow of the paper.
%Such an appendix should be printed before the Bibliography.
%
%
%-------------------------------------------------------------------------------
% If you use biblatex and either biber or bibtex to process the bibliography
% just say \printbibliography here
\printbibliography
% If you want to use the traditional BibTeX you need to use the syntax below.
%\bibliographystyle{bibtex/bst/atlasBibStyleWoTitle}
%\bibliography{main,bibtex/bib/ATLAS}
%-------------------------------------------------------------------------------
\newpage % ATLAS Collaboration author list
% Data extracted on 08-Jun-2016 for paper reference TOPQ-2015-09
\begin{flushleft}
{\Large The ATLAS Collaboration}

\bigskip

M.~Aaboud$^\textrm{\scriptsize 136d}$,
G.~Aad$^\textrm{\scriptsize 87}$,
B.~Abbott$^\textrm{\scriptsize 114}$,
J.~Abdallah$^\textrm{\scriptsize 65}$,
O.~Abdinov$^\textrm{\scriptsize 12}$,
B.~Abeloos$^\textrm{\scriptsize 118}$,
R.~Aben$^\textrm{\scriptsize 108}$,
O.S.~AbouZeid$^\textrm{\scriptsize 138}$,
N.L.~Abraham$^\textrm{\scriptsize 150}$,
H.~Abramowicz$^\textrm{\scriptsize 154}$
% ...
\bigskip
\\
% ...
$^{12}$ Institute of Physics, Azerbaijan Academy of Sciences, Baku, Azerbaijan\\
% ...
$^{65}$ University of Iowa, Iowa City IA, United States of America\\
% ...
$^{87}$ CPPM, Aix-Marseille Universit{\'e} and CNRS/IN2P3, Marseille, France\\
% ...
$^{108}$ Nikhef National Institute for Subatomic Physics and University of Amsterdam, Amsterdam, Netherlands\\
% ...
$^{114}$ Homer L. Dodge Department of Physics and Astronomy, University of Oklahoma, Norman OK, United States of America\\
% ...
$^{118}$ LAL, Univ. Paris-Sud, CNRS/IN2P3, Universit{\'e} Paris-Saclay, Orsay, France\\
% ...
$^{136}$ $^{(a)}$ Facult{\'e} des Sciences Ain Chock, R{\'e}seau Universitaire de Physique des Hautes Energies - Universit{\'e} Hassan II, Casablanca; $^{(b)}$ Centre National de l'Energie des Sciences Techniques Nucleaires, Rabat; $^{(c)}$ Facult{\'e} des Sciences Semlalia, Universit{\'e} Cadi Ayyad, LPHEA-Marrakech; $^{(d)}$ Facult{\'e} des Sciences, Universit{\'e} Mohamed Premier and LPTPM, Oujda; $^{(e)}$ Facult{\'e} des sciences, Universit{\'e} Mohammed V, Rabat, Morocco\\
% ...
$^{138}$ Santa Cruz Institute for Particle Physics, University of California Santa Cruz, Santa Cruz CA, United States of America\\
% ...
$^{150}$ Department of Physics and Astronomy, University of Sussex, Brighton, United Kingdom\\
% ...
$^{154}$ Raymond and Beverly Sackler School of Physics and Astronomy, Tel Aviv University, Tel Aviv, Israel\\
% ...
$^{a}$ Also at Department of Physics, King's College London, London, United Kingdom\\
$^{b}$ Also at Institute of Physics, Azerbaijan Academy of Sciences, Baku, Azerbaijan\\
$^{c}$ Also at Novosibirsk State University, Novosibirsk, Russia\\
$^{d}$ Also at TRIUMF, Vancouver BC, Canada\\
$^{e}$ Also at Department of Physics {\&} Astronomy, University of Louisville, Louisville, KY, United States of America\\
% ...
\end{flushleft}
% Created with xml2latex.py

%%-------------------------------------------------------------------------------
%% Print the list of contributors to the analysis
%% The argument gives the fraction of the text width used for the names
%%-------------------------------------------------------------------------------
%\clearpage
%\PrintAtlasContribute{0.30}
%
%%-------------------------------------------------------------------------------
%\clearpage
%\appendix
%\part*{Auxiliary material}
%\addcontentsline{toc}{part}{Auxiliary material}
%%%-------------------------------------------------------------------------------
%\input{AuxPlots}
\end{document}